\documentclass[aps,prd,reprint,superscriptaddress,nofootinbib]{revtex4-2}
\usepackage{orcidlink}
\usepackage{upgreek}
\usepackage{graphicx}
\usepackage{dcolumn}
\usepackage{bm}
\usepackage{enumitem}
\usepackage{mathtools}
\usepackage[table]{xcolor}
\usepackage{colortbl,booktabs}
\usepackage[mathlines]{lineno}
\usepackage[normalem]{ulem}
\usepackage{multirow}
\usepackage{empheq}
\usepackage{amsmath}
\usepackage{comment}
\hypersetup{
  colorlinks=true,
  linkcolor=blue,
  citecolor=blue,
  urlcolor=blue
}
\newcommand{\planck}{{\it Planck }}
\newcommand{\agora}{\textsc{Agora}}

\newcommand{\aap}{Astron. Astrophys.}
\newcommand{\aapr}{Astron. Astrophys. Rev.}
\newcommand{\apjl}{Astrophys. J. Lett.}
\newcommand{\apjs}{Astrophys. J. Suppl.}
\newcommand{\araa}{Annu. Rev. Astron. Astrophys.}
\newcommand{\jcap}{J. Cosmol. Astropart. Phys.}
\newcommand{\mnras}{Mon. Not. R. Astron. Soc.}
\newcommand{\pasp}{Publ. Astron. Soc. Pac.}

\begin{document}


\title{SPT-3G D1: Compton-$y$ maps using data from the SPT-3G and \planck surveys}


\affiliation{Kavli Institute for Particle Astrophysics and Cosmology, Stanford University, 452 Lomita Mall, Stanford, CA, 94305, USA}
\affiliation{Department of Physics, Stanford University, 382 Via Pueblo Mall, Stanford, CA, 94305, USA}
\affiliation{SLAC National Accelerator Laboratory, 2575 Sand Hill Road, Menlo Park, CA, 94025, USA}
\affiliation{California Institute of Technology, 1200 East California Boulevard., Pasadena, CA, 91125, USA}
\affiliation{Department of Physics \& Astronomy, University of California, One Shields Avenue, Davis, CA 95616, USA}
\affiliation{Center for AstroPhysical Surveys, National Center for Supercomputing Applications, Urbana, IL, 61801, USA}
\affiliation{Fermi National Accelerator Laboratory, MS209, P.O. Box 500, Batavia, IL, 60510, USA}
\affiliation{Department of Astronomy and Astrophysics, University of Chicago, 5640 South Ellis Avenue, Chicago, IL, 60637, USA}
\affiliation{Kavli Institute for Cosmological Physics, University of Chicago, 5640 South Ellis Avenue, Chicago, IL, 60637, USA}
\affiliation{School of Physics, University of Melbourne, Parkville, VIC 3010, Australia}
\affiliation{Sorbonne Universit\'e, CNRS, UMR 7095, Institut d'Astrophysique de Paris, 98 bis bd Arago, 75014 Paris, France}
\affiliation{Department of Physics and Astronomy, University of New Mexico, Albuquerque, NM, 87131, USA}
\affiliation{School of Physics and Astronomy, Cardiff University, Cardiff, CF24 3AA, UK}
\affiliation{High-Energy Physics Division, Argonne National Laboratory, 9700 South Cass Avenue, Lemont, IL, 60439, USA}
\affiliation{University Observatory, Faculty of Physics, Ludwig-Maximilians-Universit{\"a}t, Scheinerstr.~1, 81679 Munich, Germany}
\affiliation{Enrico Fermi Institute, University of Chicago, 5640 South Ellis Avenue, Chicago, IL, 60637, USA}
\affiliation{Department of Physics, University of Chicago, 5640 South Ellis Avenue, Chicago, IL, 60637, USA}
\affiliation{Universit\'e de Gen\`eve, D\'epartement de Physique Th\'eorique, 24 Quai Ansermet, CH-1211 Gen\`eve 4, Switzerland}
\affiliation{Department of Physics \& Astronomy, University of Sussex, Brighton BN1 9QH, UK}
\affiliation{National Taiwan University, No. 1, Sec. 4, Roosevelt Road, Taipei 106319, Taiwan}
\affiliation{Department of Physics, University of California, Berkeley, CA, 94720, USA}
\affiliation{Universit\'e Paris-Saclay, Universit\'e Paris Cit\'e, CEA, CNRS, AIM, 91191, Gif-sur-Yvette, France}
\affiliation{Department of Astronomy, University of Illinois Urbana-Champaign, 1002 West Green Street, Urbana, IL, 61801, USA}
\affiliation{High Energy Accelerator Research Organization (KEK), Tsukuba, Ibaraki 305-0801, Japan}
\affiliation{Department of Physics and McGill Space Institute, McGill University, 3600 Rue University, Montreal, Quebec H3A 2T8, Canada}
\affiliation{Canadian Institute for Advanced Research, CIFAR Program in Gravity and the Extreme Universe, Toronto, ON, M5G 1Z8, Canada}
\affiliation{Joseph Henry Laboratories of Physics, Jadwin Hall, Princeton University, Princeton, NJ 08544, USA}
\affiliation{Department of Astrophysical and Planetary Sciences, University of Colorado, Boulder, CO, 80309, USA}
\affiliation{Department of Astronomy, University of Science and Technology of China, Hefei 230026, China}
\affiliation{School of Astronomy and Space Science, University of Science and Technology of China, Hefei 230026}
\affiliation{Department of Physics, University of Illinois Urbana-Champaign, 1110 West Green Street, Urbana, IL, 61801, USA}
\affiliation{Department of Physics and Astronomy, University of California, Los Angeles, CA, 90095, USA}
\affiliation{Department of Physics and Astronomy, Michigan State University, East Lansing, MI 48824, USA}
\affiliation{Department of Physics and Astronomy, Northwestern University, 633 Clark St, Evanston, IL, 60208, USA}
\affiliation{CASA, Department of Astrophysical and Planetary Sciences, University of Colorado, Boulder, CO, 80309, USA }
\affiliation{Department of Physics, University of Colorado, Boulder, CO, 80309, USA}
\affiliation{Department of Astronomy, Cornell University, Ithaca, NY 14853, USA}
\affiliation{Department of Physics, Case Western Reserve University, Cleveland, OH, 44106, USA}
\affiliation{Department of Physics \& Astronomy, Box 41051, Texas Tech University, Lubbock TX 79409-1051, USA}
\affiliation{Dunlap Institute for Astronomy \& Astrophysics, University of Toronto, 50 St. George Street, Toronto, ON, M5S 3H4, Canada}
\affiliation{David A. Dunlap Department of Astronomy \& Astrophysics, University of Toronto, 50 St. George Street, Toronto, ON, M5S 3H4, Canada}
\affiliation{NSF-Simons AI Institute for the Sky (SkAI), 172 E. Chestnut St., Chicago, IL 60611, USA}
\affiliation{Center for Astrophysics \textbar{} Harvard \& Smithsonian, 60 Garden Street, Cambridge, MA, 02138, USA}
\author{A.~S.~Maniyar\,\orcidlink{0000-0002-4617-9320}}
\affiliation{Kavli Institute for Particle Astrophysics and Cosmology, Stanford University, 452 Lomita Mall, Stanford, CA, 94305, USA}
\affiliation{Department of Physics, Stanford University, 382 Via Pueblo Mall, Stanford, CA, 94305, USA}
\affiliation{SLAC National Accelerator Laboratory, 2575 Sand Hill Road, Menlo Park, CA, 94025, USA}
\author{F.~Bianchini\,\orcidlink{0000-0003-4847-3483}}
\affiliation{Kavli Institute for Particle Astrophysics and Cosmology, Stanford University, 452 Lomita Mall, Stanford, CA, 94305, USA}
\affiliation{Department of Physics, Stanford University, 382 Via Pueblo Mall, Stanford, CA, 94305, USA}
\affiliation{SLAC National Accelerator Laboratory, 2575 Sand Hill Road, Menlo Park, CA, 94025, USA}
\author{W.~L.~K.~Wu\,\orcidlink{0000-0001-5411-6920}}
\affiliation{California Institute of Technology, 1200 East California Boulevard., Pasadena, CA, 91125, USA}
\affiliation{Kavli Institute for Particle Astrophysics and Cosmology, Stanford University, 452 Lomita Mall, Stanford, CA, 94305, USA}
\affiliation{SLAC National Accelerator Laboratory, 2575 Sand Hill Road, Menlo Park, CA, 94025, USA}
\author{S.~Raghunathan\,\orcidlink{0000-0003-1405-378X}}
\affiliation{Department of Physics \& Astronomy, University of California, One Shields Avenue, Davis, CA 95616, USA}
\affiliation{Center for AstroPhysical Surveys, National Center for Supercomputing Applications, Urbana, IL, 61801, USA}
\author{A.~J.~Anderson\,\orcidlink{0000-0002-4435-4623}}
\affiliation{Fermi National Accelerator Laboratory, MS209, P.O. Box 500, Batavia, IL, 60510, USA}
\affiliation{Kavli Institute for Cosmological Physics, University of Chicago, 5640 South Ellis Avenue, Chicago, IL, 60637, USA}
\affiliation{Department of Astronomy and Astrophysics, University of Chicago, 5640 South Ellis Avenue, Chicago, IL, 60637, USA}
\author{B.~Ansarinejad}
\affiliation{School of Physics, University of Melbourne, Parkville, VIC 3010, Australia}
\author{M.~Archipley\,\orcidlink{0000-0002-0517-9842}}
\affiliation{Department of Astronomy and Astrophysics, University of Chicago, 5640 South Ellis Avenue, Chicago, IL, 60637, USA}
\affiliation{Kavli Institute for Cosmological Physics, University of Chicago, 5640 South Ellis Avenue, Chicago, IL, 60637, USA}
\author{L.~Balkenhol\,\orcidlink{0000-0001-6899-1873}}
\affiliation{Sorbonne Universit\'e, CNRS, UMR 7095, Institut d'Astrophysique de Paris, 98 bis bd Arago, 75014 Paris, France}
\author{D.~R.~Barron\,\orcidlink{0000-0002-1623-5651}}
\affiliation{Department of Physics and Astronomy, University of New Mexico, Albuquerque, NM, 87131, USA}
\author{P.~S.~Barry\,\orcidlink{0000-0001-9103-9354}}
\affiliation{School of Physics and Astronomy, Cardiff University, Cardiff, CF24 3AA, UK}
% ...
\collaboration{SPT-3G Collaboration}
\noaffiliation


\begin{abstract}
We present thermal Sunyaev-Zel'dovich (tSZ) Compton-$y$ parameter maps constructed from two years (2019-2020) of observations with the South Pole Telescope (SPT) third-generation camera, SPT-3G, combined with data from the \planck satellite. Using a linear combination (LC) pipeline, we obtain a suite of reconstructions that explore different trade-offs between statistical sensitivity and suppression of astrophysical contaminants, including minimum-variance, CMB-deprojected, and CIB-deprojected $y$-maps. We validate these maps through different statistical techniques such as auto- and cross-power spectra with large-scale structure tracers as well as stacking on cluster locations. These tests are used to understand the balance between noise and astrophysical foreground residuals (such as the CIB) in combination with the recovery of the tSZ signal for different maps. For example, results from stacking at the location of clusters confirm the robustness of the recovered tSZ signal over the $\sim 1500\: {\rm deg}^2$ SPT-3G survey field used in this analysis. The high-resolution and low-noise maps produced here provide an important cosmological tool for future studies, including measurements of the Compton-$y$ map power spectrum, cross-correlations with other tracers of the large-scale structure, detailed modeling of cluster pressure profiles, and study of the thermodynamic state of the baryons in the Universe. 
\end{abstract}

\maketitle


\section{Introduction}\label{sec:intro}

Modern experiments such as \planck \cite{planck2011, plancklegacy2020}, the Atacama Cosmology Telescope (ACT) \cite{Swetz2011, act2020}, and the South Pole Telescope (SPT) \cite{Carlstrom2011, Quan2026} have mapped the cosmic microwave background (CMB) sky with exquisite sensitivity and angular resolution in temperature as well as polarization. These observations of the CMB have transformed our understanding of the cosmology of the Universe \cite{planckcosmo2020, muse2025, actdr4cosmo, Camphuis2025}. Alongside the dominant relic radiation from the early Universe, the CMB maps are imprinted with astrophysical foregrounds such as Galactic dust \cite{Boulanger1996}, synchrotron \cite{Zotti2010}, and free-free emission \cite{Smoot1998}, as well as extragalactic emissions like the cosmic infrared background (CIB) \cite{Puget1996} and radio sources \cite{Zotti2010}. Moreover, secondary anisotropies arise from the interaction of CMB photons with matter at later times, such as gravitational lensing, the integrated Sachs-Wolfe effect, and the Sunyaev–Zel’dovich (SZ) effects, providing a wealth of additional information (see e.g.~\citep{Aghanim2008, Bianchini2025} for a review of the CMB secondary anisotropies). Extracting each of these secondary contributions requires careful component separation analysis.

The SZ effects, originating from the scattering of the CMB photons off free electrons, are excellent tracers of large-scale structure at intermediate redshifts \cite{Sunyaev1970, Sunyaev1972}. The thermal SZ (tSZ) effect arises from the inverse Compton scattering of CMB photons by hot electrons in galaxy clusters and the intergalactic medium. This process distorts the CMB spectrum in a frequency-dependent manner, producing a characteristic decrement in intensity below $\sim 217$ GHz and an increment above it \cite{Sunyaev1970, Sunyaev1972}. In contrast, the kinetic SZ (kSZ) effect is generated by the Doppler shift from bulk motions of ionized gas, which preserves the CMB blackbody spectrum. These signals carry complementary cosmological information. While the tSZ effect is particularly sensitive to the thermal pressure of baryons and provides a powerful probe of the thermodynamic history of the Universe, the kSZ effect encodes information about the density of baryons and the large-scale velocity fields (see \cite{Carlstrom2002} for a review).

In this paper, we focus on isolating one such component from the CMB maps: the tSZ effect. Over the last two decades, the tSZ effect has emerged as a central tool for cosmology and astrophysics. The tSZ effect enables compilation of mass-limited cluster samples out to high redshifts \cite{Bleem2020, Hilton2021}. Cluster-based measurements of the tSZ signal have been used to constrain cosmological parameters through cluster counts and scaling relations \cite{planckclusters2016, Bocquet2019}. The power spectrum and the bispectrum of the tSZ effect also provide valuable cosmological insights \cite{Crawford2014, Horowitz2017}. At the same time, map-level reconstructions of the tSZ signal provide a more complete, unbiased picture of hot baryons, capturing not only the contribution from massive clusters but also diffuse gas in groups, filaments, and the outskirts of halos \cite{Battaglia2012, Ruppin2021}. These maps enable statistical studies such as characterization of the elusive `missing baryons' \cite{Graaff2019, Tanimura2019}. They also provide inputs for cross-correlation analyses with weak lensing \cite{Pandey2025, Waerbeke2014}, galaxy clustering \cite{Koukoufilippas2020}, and CMB lensing \cite{Hill2014}, thereby offering unique insights into the interplay between dark matter and baryons across cosmic time.

Given its scientific value, considerable effort has gone into developing techniques for extracting tSZ maps from multifrequency CMB data (e.g. among other examples, \cite{Remazeilles2011a, Remazeilles2011b, Hurier2013, Abylkairov2021}). The primary challenge is separating the tSZ signal from other sky components with overlapping spatial and spectral characteristics, most notably the CMB itself and the CIB. A widely used class of methods is based on linear combinations of multifrequency maps that exploit the distinct spectral dependence of the tSZ effect. Internal linear combination (ILC) approach \cite{Tegmark1998}, for example, is one such method which was first applied to the Cosmic Background Explorer (\textit{COBE}) \cite{cobe1990} data. This method can be used to construct weighted combinations of maps that preserve the tSZ signal, while optionally nulling or suppressing the CMB and foregrounds such as the CIB. A number of groups have advanced these efforts. For instance, the modified ILC algorithm (MILCA) and needlet ILC (NILC) pipelines exploited generalized ILC techniques and applied them to \planck data \cite{plancknilc2016,Chandran2023, McCarthy2024}. Unlike standard ILCs, these methods operate locally (in patches or a wavelet basis), allowing them to adapt to spatially-varying foregrounds and noise. These developments also include constrained ILC (cILC) methods, which impose explicit spectral constraints to preserve the tSZ signal while nulling selected contaminants such as the CMB \cite{Remazeilles2011a}. More recently, sensitive high-resolution data from ACT and SPT have enabled deeper reconstructions on smaller regions of the sky, complementing \planck with finer angular information \cite{Madhavacheril2020, Bleem2022}. Some of these works have combined the sensitive ground based data from ACT/SPT with \planck data \cite{Aghanim2019, Bleem2022, Coulton2024} to produce maps that cover a wide range of scales. 
These works have demonstrated the power of combining multifrequency data sets and developing tailored linear combination methods to get high-fidelity tSZ reconstructions.

In this paper, we extend these efforts by constructing a new tSZ Compton-$y$ parameter map that combines data from SPT and the \planck satellite. The main survey of the SPT-3G experiment provides high-resolution millimeter-wave maps over $\sim 1500 \:\mathrm{deg}^2$ of the southern sky, with deep coverage in the 95, 150, and 220 GHz bands. Meanwhile, \planck delivers full-sky coverage at lower angular resolution but across a wider frequency range. By combining the 2019–2020 two-year SPT-3G main data set taken during the austral winters (which we refer to as SPT- 3G D1) with {\it Planck}, we leverage the complementary strengths of the two experiments.
Our approach is based on a two-dimensional harmonic-space linear combination analysis. We construct weights in harmonic space that preserve the spectral shape of the tSZ effect while suppressing contamination from other components. We construct several versions of the tSZ map smoothed with a $1'.4$ beam, including minimum-variance (MV), CMB-deprojected, and CIB-deprojected variants, each tailored for different scientific applications.

This paper is organized as follows. In Section~\ref{sec:data}, we describe the data sets and simulations used to construct the maps. Section~\ref{sec:ilc} describes the methodology of the harmonic-space ILC that we use with our detailed results described in Section~\ref{sec:results}. Finally, we summarize our results and present the conclusions in Section~\ref{sec:conclusion}. All the public data products in this paper will be made available on the public SPT website.\footnote{\url{https://pole.uchicago.edu/public/data/}}
\section{Data and simulations}
\label{sec:data}
The tSZ maps presented here are constructed using the data from the $1500 \: {\rm deg}^2$ SPT-3G survey and the \planck all-sky survey PR3 data release.\footnote{\url{https://pla.esac.esa.int/\#maps}} 
The lower \planck noise at multipoles $\ell \lesssim 1500$ with enhanced frequency coverage and higher resolution SPT-3G data with low noise on small scales are extremely complementary to each other, and we make use of these features in this work. We use \agora ~simulations \cite{Omori2024} to estimate the contribution of different extra-galactic components to our analysis and to validate our pipeline. In this section, we summarize these data and simulation products as relevant to the construction of the tSZ maps. 

\subsection{SPT-3G D1}
\label{subs:spt3gdata}
SPT is a 10-meter aperture telescope located at the Amundsen-Scott South Pole Station close to the geographical South Pole in Antarctica \cite{Carlstrom2011}. Its third-generation camera, SPT-3G, has been in operation since 2017, following the earlier SPT-SZ and SPTpol instruments. SPT-3G features a large focal plane with $\sim 16000$ superconducting transition-edge sensor bolometers, which is an order of magnitude upgrade from the previous cameras. This enables sensitive measurements of both the intensity and polarization of the CMB. This high detector count observing simultaneously in three frequency bands centered at 95, 150, and 220 GHz, provides the multifrequency leverage and exceptional sensitivity needed for producing high-fidelity tSZ (Compton-$y$) maps.

\subsubsection{Map-making}
\label{subsub:sptmapmaking}
We use SPT-3G data from the austral winter observing seasons of 2019 and 2020. These data come from the SPT-3G Main field which covers approximately 4\% of the full sky ($\sim 1500\: {\rm deg}^2$) with a footprint that spans declinations from $-42^\circ$ to $-70^\circ$ and right ascensions from 20h40m0s to 3h30m0s.
We employ sky maps produced by the SPT-3G collaboration with a data-processing and map-making pipeline that is outlined in \cite{Quan2026, Camphuis2025}. In brief, co-added temperature 
maps for each observing band (95, 150, and 220 GHz) are generated from time-ordered data from the Main field which is subdivided into four sub-fields. Each sub-field covers only one-fourth of the entire range of declination  of the Main field while spanning the entire range of right ascension. This is done in order to avoid detector nonlinearity from large changes in atmospheric loading with elevation. 

The time-ordered data are initially processed using both low-pass and high-pass filters to suppress unwanted noise. Low-pass filtering, implemented in Fourier space, attenuates high-frequency noise and mitigates aliasing. High-pass filtering is performed by fitting the timestream of every detector to a combination of low-order polynomials and low-frequency sinusoids, effectively removing large-scale instrumental and atmospheric noise. Subsequently, the maps are constructed with these timestreams alongside inverse-variance `weight maps' that quantify per-pixel noise levels. There are also a large number of point sources present in our maps, which, if included in the parametric fit used to high-pass filter the data, result in large scan-direction features near the locations of those sources. In order to avoid these features, we remove the point source (above 6 mJy at 150 GHz) regions from the parametric fit \cite{Camphuis2025, Quan2026}. As a result of the high-pass filtering step, modes below $\ell < 300$ along the scan direction are suppressed and there is little information in the maps below these multipoles. We discuss the effect of these missing modes in Section~\ref{subsub:mmodesptplanck}. We also produce two splits of maps by dividing individual observations into two parts and adding each half together, obtaining maps with same underlying signal but uncorrelated noise \cite{Quan2026}. 

Finally, the maps are made using \texttt{HEALPix}\footnote{\url{https://healpix.sourceforge.io/}} \cite{healpix2005} pixelization with $N_{\rm side}$ parameter of 8192, corresponding to an angular resolution of approximately $0'.4$. We take advantage of the existing tools optimized for efficient computation of spherical harmonic transforms (SHTs) and power spectra from HEALPix-formatted maps and perform this analysis using full-sky SHTs.

\subsubsection{Noise Power Spectra}
A detailed description of the construction of the noise maps is provided in \cite{Camphuis2025, Quan2026}. The 500 noise realizations are generated by using a `sign-flip' method to isolate noise fluctuations. First, at each frequency the full-depth map is subtracted from the map of each individual observation to remove the sky signals. Then, to generate one noise realization, the full set of individual observations is randomly split into two parts of roughly equal signal-to-noise. After multiplying all the individual-observation maps in one part by $-1$, the maps from the two parts are added together with the inverse-variance weighting.
Because the sign flips cancel out signal contributions but retain noise variance, the resulting maps provide realistic samples of the instrument and atmospheric noise. Such realizations are then used to assess noise power spectra.
The white noise amplitudes at 95, 150, and 220 GHz channels come out to be 5, 4, and 16 $\mu {\rm K}\textnormal{-}{\rm arcmin}$ respectively.

\subsubsection{Calibration and Beams}
Absolute calibration factors are independently determined for each sub-field by cross-correlating SPT-3G maps with \planck PR3 data observations, which provide a reliable absolute reference through measurements of the CMB dipole amplitude. More specifically, we calibrate SPT-3G 150 GHz sub-fields with the \planck 143 GHz maps. This is done using a cross-spectrum approach taking the ratio of the cross-power spectrum between 150 and 143 GHz maps with the cross-spectrum of two halves of SPT 150 GHz map. We then calculate the inverse-variance weighted average of this ratio in the range $800 < \ell < 1200$ to get the calibration factor for that field. Then we perform internal calibration for our 95 and 220 GHz maps using the calibrated 150 GHz map. These calibration factors are then applied to the respective sub-fields, which are subsequently co-added to produce the final full-field maps. All the details for this procedure, including the beam and map level corrections, are described in detail in \cite{Quan2026}.

The 10-meter primary mirror of SPT enables its high angular resolution. The point spread function of the instrument, commonly referred to as the beam, which is determined by the full optical system, suppresses small-scale modes in the observed signal.
Accurate characterization of this beam is therefore essential to recover the true underlying sky signal.
For SPT, the fiducial temperature beam is constructed using observations of both planets and bright astrophysical sources, primarily active galactic nuclei (AGNs). These two classes of sources are used to characterize different angular scales of the beam (Huang \textit{et al.}, in preparation).
Planetary observations, due to the high brightness of these objects, allow for precise measurement of the beam profile at large angular scales up to several tens of arcminutes. However, because planets are extremely bright, they tend to saturate the detectors in the central (inner core) region of the beam. To overcome this, AGNs being more distant (lower flux levels), are used to characterize the beam at smaller angular scales, particularly in the inner core region.
These two sets of measurements are then combined to produce a composite beam profile that spans from the inner core to the outer regions of the beam (Huang \textit{et al.}, in preparation). These beams correspond to effective Gaussian beams of full width at half maximum (FWHM) = $1'.6$, $1'.2$, and $1'.0$ respectively at 95, 150, and 220 GHz.


\subsection{\planck data}
\label{subs:planckdata}
\planck was a European Space Agency (ESA) satellite launched in 2009 to make full-sky maps of the CMB anisotropies \cite{planck2011}. This mission produced full-sky maps of the sky from 30-857 GHz in nine frequency bands. In this analysis, we make use of the data from \planck PR3. 
In particular, we use high frequency instrument (HFI) data maps at 100, 143, 217, and 353 GHz.\footnote{\url{https://pla.esac.esa.int/pla/}} 
We use both full- and half-mission maps and even-odd ring maps. Using simulations, we checked that adding \planck 545 and 857 GHz channels did not improve our results in terms of noise or foreground residuals, and so we do not include them in our analysis.

\subsubsection{Maps, Calibration, Beams, and Noise Power Spectra}
\label{subsub:pladata}
The detailed process to convert the \planck time-ordered data into maps for each channel is described in \cite{planckmapmk2014} and updated in \cite{planckhfidata2016}. These maps are provided in \texttt{HEALPix} format with $N_{\rm side} = 2048$, which corresponds to 1'.7 pixels. 

The 100, 143, 217, and 353 GHz HFI maps are calibrated using the time-varying CMB dipole \cite{planckhfidata2016}. As a result, the absolute calibration of these full-mission HFI maps is determined to 0.09, 0.07, 0.16, and 0.78\% respectively at 100, 143, 217, and 353 GHz. The \planck collaboration also provides an effective beam which accounts for telescope optics, data processing convolved with map pixelization, and survey scan strategy, and acts as the transfer function as explained in \cite{planckbeams2016}. These beams correspond to effective Gaussian beams of FWHM = 9'.66, 7'.27, 5'.01, and 4'.86 respectively at 100, 143, 217, and 353 GHz. 

We use the HFI half-mission maps at 100, 143, 217, and 353 GHz to get the noise power spectra. Compared to the full-sky average, the noise power spectra levels in these maps are up to $\sim 30\%$ lower in the regions near the south ecliptic pole in SPT-3G footprint because the \planck observation strategy results in deeper coverage in this region. As mentioned before, the \planck data mostly serve to fill in the large angular scale modes that are missing from SPT maps, and this lower noise level further helps in making the resulting tSZ maps less noisy on large angular scales. 

\subsection{\agora ~simulations}
In order to validate our pipeline, as well as to get the expected power spectra of different foreground components contributing to the maps, we use \agora ~simulations \cite{Omori2024}. \agora ~is a simulation suite of extragalactic foregrounds, including tSZ, kSZ, CIB, and radio sources. These maps are further lensed according to the underlying matter density field, producing more realistic sky realizations. These raw sky realizations are then convolved with the actual SPT-3G and \planck bandpasses for their respective frequency channels to mimic the sky as observed by these telescopes barring the instrumental and atmospheric effects. For details on how different extragalactic foregrounds are simulated in the \agora ~simulations, we refer to \cite{Omori2024}.

\subsection{Bright point sources}
Due to the sensitive nature of our maps, a number of point sources are present in them. We apply a point source mask to both SPT-3G and \planck maps, based on the locations  of point sources detected above a threshold of 6 mJy at 150 GHz in SPT-3G data. The masking radii were determined based on the radius at which the beam-convolved point source profile falls below a signal-to-noise ratio of 1.
After masking the point sources, we perform an `inpainting' procedure at their locations. This process involves replacing the pixels where the sources are localized with a Gaussian constrained realization of the CMB informed by the surrounding pixels. This process avoids the problem of mode-coupling caused by the point source holes in the maps which cause correlations between the small-scale and large-scale modes. 
Contrary to some other SPT analyses using the same data (e.g. \cite{Camphuis2025, Quan2026}), we do not mask galaxy clusters since they are expected to contribute to the tSZ signal. A similar emissive source mask using a flux cut of 6 mJy at 150 GHz is also applied to our simulated SPT and \planck maps from the \agora ~simulations.

\section{Methodology}
\label{sec:ilc}
In this section, we start by discussing the ILC algorithm in general and our implementation for this work. A similar derivation of the ILC, `constrained' ILC, and also `Needlet ILC' (NILC), is presented in \cite{McCarthy2024}.
\subsection{ILC and constrained ILC (cILC)}
We model the observed temperature (or intensity) in frequency channel $i$ in a given direction on the sky $\hat{n}$ as the sum of a target component with known frequency dependence and other contaminating contributions:
\begin{equation}
T_i(\hat{n}) = a_i\,s(\hat{n}) + n_i(\hat{n}) \,,
\end{equation}
where $s(\hat{n})$ denotes the sky signal of interest (the Compton-$y$ field in our case), $a_i$ is the known spectral energy distribution (SED) of that signal in channel $i$, and $n_i(\hat{n})$ contains all other contributions (primary CMB, Galactic and extragalactic foregrounds, instrumental noise, etc.). Any linear combination
\begin{equation}
\tilde{s}(\hat{n}) \equiv \sum_i w_i(\mathcal{D})\,T_i(\hat{n})
\end{equation}
that enforces the unit response constraint
\begin{equation}
\sum_i w_i(\mathcal{D})\,a_i = 1
\end{equation}
is unbiased to the target signal $s(\hat{n})$. Here $\mathcal{D}$ denotes the domain in which the weights are computed (e.g. a real-space patch, a spherical harmonic multipole $\ell$, or the $(\ell,m)$ harmonic cell). Given the above constraints, for a minimum-variance reconstruction in domain $\mathcal{D}$, the weights take the form
\begin{equation}
{\bf w}(\mathcal{D}) \;=\; \frac{{\bf C}(\mathcal{D})^{-1}{\bf a}}{{\bf a}^T {\bf C}(\mathcal{D})^{-1}{\bf a}} \,,
\label{eq:ilc_weights_general}
\end{equation}
where ${\bf C}(\mathcal{D})$ is the frequency-frequency covariance matrix estimated in the domain $\mathcal{D}$, and ${\bf a}$ and ${\bf w}$ are the SED ($a_i$) and weight ($w_i$) vectors respectively. 
\bigskip

\noindent {\bf Constrained ILC:} In many situations, we wish to not only preserve the response to the target signal $s(\hat{n})$ with SED ${\bf a}$, but also to enforce zero response to one or more contaminating components with known SEDs. This leads to the cILC (or `deprojected' ILC). Suppose we want to remove $N_c$ contaminants, each described by a frequency vector ${\bf b}^{(k)}$ for $k=1,\dots,N_c$. Collecting these into a constraint matrix
\begin{equation}
{\bf F} \;=\; \big[\, {\bf a}, {\bf b}^{(1)}, {\bf b}^{(2)}, \dots, {\bf b}^{(N_c)} \,\big] \,,
\end{equation}
we require that the weight vector ${\bf w}(\mathcal{D})$ satisfies
\begin{equation}
{\bf w}(\mathcal{D})^T {\bf F} = ({\bf 1}, {\bf 0}, {\bf 0}, \dots, {\bf 0}) \,.
\label{eq:constraints}
\end{equation}

The minimum-variance solution is obtained by introducing Lagrange multipliers subject to these linear constraints of Eq.~\ref{eq:constraints}. The result is
\begin{equation}
{\bf w}(\mathcal{D}) \;=\; {\bf C}(\mathcal{D})^{-1}{\bf F}\,\big[\, {\bf F}^T {\bf C}(\mathcal{D})^{-1}{\bf F} \,\big]^{-1} {\bf e}_1 \,,
\label{eq:constrained_ilc}
\end{equation}
where ${\bf e}_1 = (1,0,0,\dots,0)^T$ enforces unit response to the target component and zero response to all contaminants. 

Compared to the standard ILC, the cILC typically yields higher variance, since additional constraints reduce the effective number of degrees of freedom available to minimize the noise. Nevertheless, it is a powerful technique when the contaminant SEDs are well understood, as it can strongly suppress leakage from specific foregrounds such as the primary CMB or the CIB.

\subsubsection{Harmonic-space ILC and Our Implementation}
\label{sec:harmonic_ilc}

A harmonic-space ILC traditionally computes an $\ell$-dependent covariance,
$C_{ij}(\ell)$, by averaging harmonic coefficients over $m$ (or over an $\ell$-bin), and then forms $\ell$-dependent weights ${\bf w}_\ell$ that are applied to each spherical-harmonic coefficient $T_{i,\ell m}$. Concretely, the usual harmonic ILC reconstructs the target multipoles as
\begin{equation}
\tilde{s}_{\ell m} \;=\; \sum_i w_{i, \ell}\,T_{i,\ell m} \;,\qquad
{\bf w}_\ell = \frac{{\bf C}_\ell^{-1}{\bf a}}{{\bf a}^T {\bf C}_\ell^{-1}{\bf a}} \,,
\end{equation}
with ${\bf C}_\ell$ being the covariance matrix.

In this work we compute 2D weights ${\bf w}_{\ell m}$ and apply them per harmonic mode:
\begin{equation}
\tilde{s}_{\ell m} \;=\; \sum_i w_{i, \ell m}\,T_{i,\ell m} \,,
\label{eq:lm_weights}
\end{equation}
with the $(\ell,m)$-dependent minimum-variance weights given by
\begin{equation}
{\bf w}_{\ell m} \;=\; \frac{{\bf C}_{\ell m}^{-1}{\bf a}}{{\bf a}^T {\bf C}_{\ell m}^{-1}{\bf a}} \,,
\label{eq:wlm}
\end{equation}
where
$C_{ij}(\ell,m) \;\equiv\; \langle T_{i,\ell m}\,T_{j,\ell m}^* \rangle$ (or an appropriately smoothed estimator of this quantity).


The $(\ell,m)$-dependence can be important. Letting the weights depend on both $\ell$ and $m$ permits the ILC to adapt to anisotropic noise and foreground structure that breaks statistical isotropy (for example, atmospheric noise or localized foregrounds). In such situations the $m$-averaged covariance ${\bf C}_\ell$ can mask important anisotropic contamination, whereas ${\bf C}_{\ell m}$ retains that information and allows the weights to suppress contamination on a mode-by-mode basis. 

\subsubsection{Covariance Matrix Construction}
SPT noise is anisotropic in nature on the sky. This effect arises from a combination of SPT scan strategy and low-frequency noise contributions from the atmosphere and other instrumental sources. Specifically, modes that vary rapidly along the scan direction are less affected by the low-frequency noise (less noisy), while slowly varying modes along the same direction tend to be more contaminated (more details in \cite{Quan2026}). As a result, using the full anisotropic ${\bf C}_{\ell m}$ covariance matrix is crucial for our work. 

\begin{figure}
	\includegraphics[width=\columnwidth]{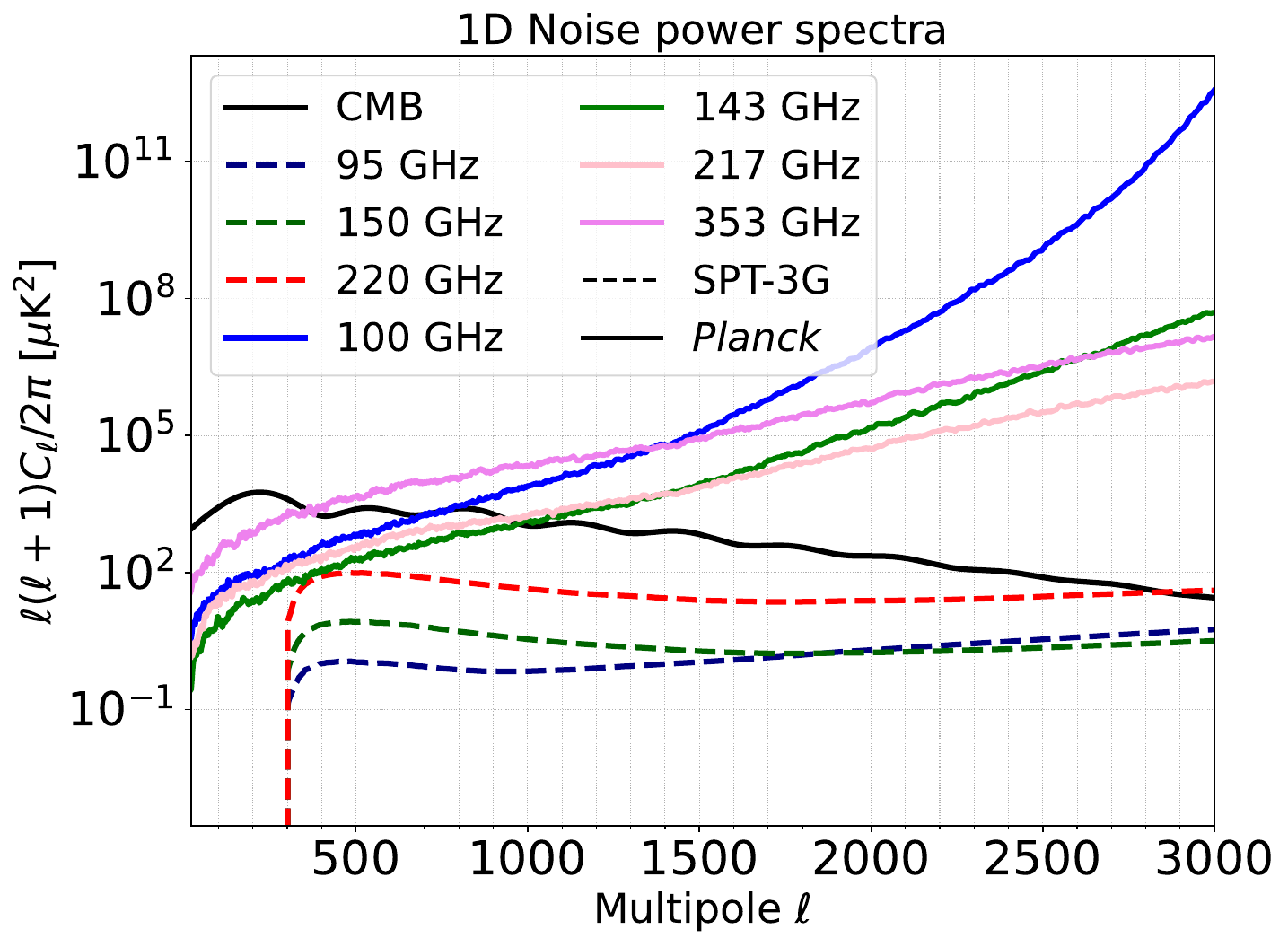}
    \caption{Beam-deconvolved 1D noise power spectra for temperature data for different \planck (solid lines) and SPT (dashed lines) frequency channels. The \planck noise curves are from within the SPT-3G patch on the sky. As pointed out in Sec.~\ref{subsub:sptmapmaking}, due to the high-pass filtering step during our map-making, SPT maps lose modes below $\ell < 300$. The black curve shows the expected CMB power spectrum using the 2018 best-fit \planck cosmology for comparison. }
    \label{fig:noisecls}
\end{figure}

As shown in Fig.~\ref{fig:noisecls}, we observe that on small-scales ($\sim \ell > 1500$) the noise power spectra in the \planck frequency channels are much higher than those in SPT. As a result, we do not expect the \planck data to add any extra information to the resulting tSZ maps. 
We therefore explicitly set all the \planck auto- and cross-power spectra in the covariance matrix to infinity after $\ell = 3000$. 

We neglect the contribution of galactic dust to the covariance matrix because its power is negligible compared to other sky components and instrumental noise at the angular scales ($\ell > 1000$) relevant for tSZ science.
At the same time, we expect all the extragalactic components (CMB, CIB, tSZ, kSZ, radio) to be isotropic across our field. As a result, for their contributions to the covariance matrix, we have ${\bf C}_{\ell m} = {\bf C}_{\ell}$. While our estimation for the noise power spectra comes directly from the data, power spectra for extragalactic components come from \agora~simulations \cite{Omori2024}. For each component, we take the full-sky \planck or SPT band-passed \agora~map, compute its one-dimensional angular power spectrum ${\bf C}_{\ell}$, and assign that value to every $m$ cell for a given $\ell$. 

It is well known that because the covariance matrix for the ILC analysis is constructed from the data, the component separated maps can be biased. This is known as the ILC bias, which arises from random correlations between the signal and the noise \cite{Delabrouille2009, Tegmark2000}. In modes where the signal and noise partially cancel, the resulting variance is reduced and as a result the ILC algorithm tends to up-weight these modes, leading to an overall suppression of the true signal. Since we mostly rely on simulations for estimating the covariance matrix, we avoid this ILC bias. At the same time, this also means that our analysis is strictly not an ILC analysis, and is a linear combination (LC) analysis since we do not use the `internal' data to get the covariance matrix. The same approach has been used in other Compton-$y$ maps from SPT collaboration (e.g. \cite{Bleem2022}).

\subsubsection{Frequency Response Functions}
\label{subsub:freqresp}
Our LC algorithm, in addition to the covariance matrix, requires a SED response function describing the contribution of the required component at each of the frequency channels. Our modeling for these contributions is as follows:
\begin{itemize}
    \item For the CMB and kSZ, we assume that all our input maps are calibrated such that they are in the thermodynamic units i.e., they have a unit response to the CMB fluctuations. Thus, the frequency response function in Eq.~\ref{eq:ilc_weights_general} is $a^{\rm CMB}_i = 1$.
    
    \item The tSZ is a spectral distortion of the CMB and its frequency response for a given frequency $\nu$ is given as 
        \begin{equation}
        \label{eq:ysed}
            a^{\rm tSZ}(x) = \left(x \frac{e^x + 1}{e^x - 1} - 4 \right) (1 + \delta_{\rm rc})\, ,
        \end{equation}
    where $x = {h\nu}/{k_{\rm B} T_{\rm CMB}}$ with $k_{\rm B}$ being the Boltzmann constant, $T_{\rm CMB} = 2.7260 \ {\rm K}$ is the CMB temperature, $h$ is Planck's constant, and ${\delta_{\rm rc}}$ encompasses the terms dependent on the temperature of the gas. In general, $a^{\rm tSZ}(x)$ depends very weakly on the temperature of the electrons in the gas that scatter the CMB \cite{Itoh1998, Challinor1998}. This temperature dependent term represents a relativistic correction and is important for highly energetic relativistic electrons i.e., when $k_B T_e \sim m_e c^2$ where $T_e$ and $m_e$ are the electron temperature and mass, and $c$ is the speed of light (see e.g. \cite{Chluba2012} for the full calculation). With the current sensitivity and improved modeling techniques, these relativistic effects have been detected (e.g. \cite{Coulton2024b, Remazeilles2025}). In this analysis, we neglect these small corrections, as a dedicated analysis of this signal is beyond the scope of this paper. 
    
    \item The CIB is one of the major extragalactic foregrounds and a contaminant for the tSZ maps. We model the frequency response of the CIB as a modified blackbody given as 
        \begin{equation}
            a^{\rm CIB}(\nu) = \frac{\nu^{3+\beta_d}}{{\rm exp} \frac{h\nu}{k_{\rm B} T_d} - 1} \left( \left.\frac{dB(\nu, T)}{dT}\right\vert_{T=T_{\rm CMB}}\right)^{-1},
        \end{equation}
        where $B(\nu, T)$ is the blackbody spectrum for a temperature $T$, and $T_d$ and $\beta_d$ denote the extragalactic dust temperature and emissivity index respectively. Several different types of star-forming galaxies at different redshifts contribute to the CIB. As a result, the SED of the CIB is quite complex and this modified blackbody formulation is a simple approximation. Furthermore, available simulations like \agora~ may not perfectly represent the real CIB sky. Over the range of frequencies probed here, $\beta_d$ and $T_d$ are fairly degenerate \citep[e.g.][]{McCarthy2024}. All of this motivates us to use a `hybrid' approach combining simulations and data for parameter estimation.  We run a blind two parameter grid search over different values of $\beta_d$ and $T_d$ on \agora~simulations to minimize the resulting CIB and overall foreground and noise residuals in the range $500 < \ell < 5000$ following \cite{Raghunathan2023}. This test favors a value of $\beta_d = 1.8$ and $T_d=10.0 \ {\rm K}$. However, we found that the resulting CIB-deprojected Compton-$y$ maps when cross-correlated with the CIB maps at 545 GHz from \cite{Lenz2019} resulted in a value consistent with zero or negative at $\ell < 2000$ (see Sec.~\ref{subsub:cibcross}). We nominally expect a positive CIB-$y$ correlation between the two maps. The choice of $\beta_d = 2.0$ and $T_d=10.0 \, {\rm K}$ results in slightly higher CIB residuals but have similar level of total residuals in simulations compared to the $\beta_d = 1.8$ and $T_d=10.0 \, {\rm K}$ case. Conversely, this choice resulted in a positive correlation with the CIB 545 and 857 GHz maps. As a result, going forward, we use $\beta_d = 2.0$ and $T_d=10.0 \, {\rm K}$ as our default values for CIB-deprojected Compton-$y$ maps based on this physical behavior of the data. 
        We prioritize this behavior over a strict minimization within an imperfect model and simulation. This tension between the simulation-based and data-driven results highlights the challenge of CIB deprojection. As we will discuss in the subsequent sections, this choice has implications for the CIB residual levels.

\end{itemize}

\subsubsection{Missing Modes From Maps}
\label{subsub:mmodesptplanck}
We point out in Sec.~\ref{subsub:sptmapmaking} that due to certain data filtering and SPT map-making strategies, certain sets of modes are lost from SPT-3G maps. More specifically, modes $m \lesssim 220$ are not completely recovered at all multipoles and modes $\ell < 300$ are lost. 

In contrast, \planck data do not suffer from this filtering and retain information in these modes. Therefore, the LC algorithm relies entirely on \planck data to reconstruct this region of Fourier space. However, on small physical scales (high $\ell$ values), the finite angular resolution of \planck leads to significant beam suppression. Correcting for this suppression requires a large multiplicative factor, which also amplifies the corresponding instrumental noise. Consequently, the region of Fourier space characterized by high-$\ell$ and low-$m$ becomes significantly noisier than average and there is effectively no information present.

We use an approach similar to the one used by \cite{Bleem2022} to mitigate this excess noise. We apply a harmonic-space filter to the output Compton-$y$ maps. This filter is a low-pass filter in the $\ell$ space for $m \leq 220$. It is set to one at low-$\ell$ values and then smoothly rolls off to zero at high-$\ell$ values, thereby smoothly suppressing the noise at high-$\ell$ values. The parameters for the low-pass filter are chosen such that the noise level within this $m \lesssim 220$ region is approximately equal to that in the regions outside of it such that there are no abrupt jumps in the power spectra.

We include this filter as part of our data release. This can be used as an effective transfer function or harmonic-space beam for subsequent power spectrum or correlation function analyses. 

\begin{figure*}
\centering
    \includegraphics[width=1.\textwidth]{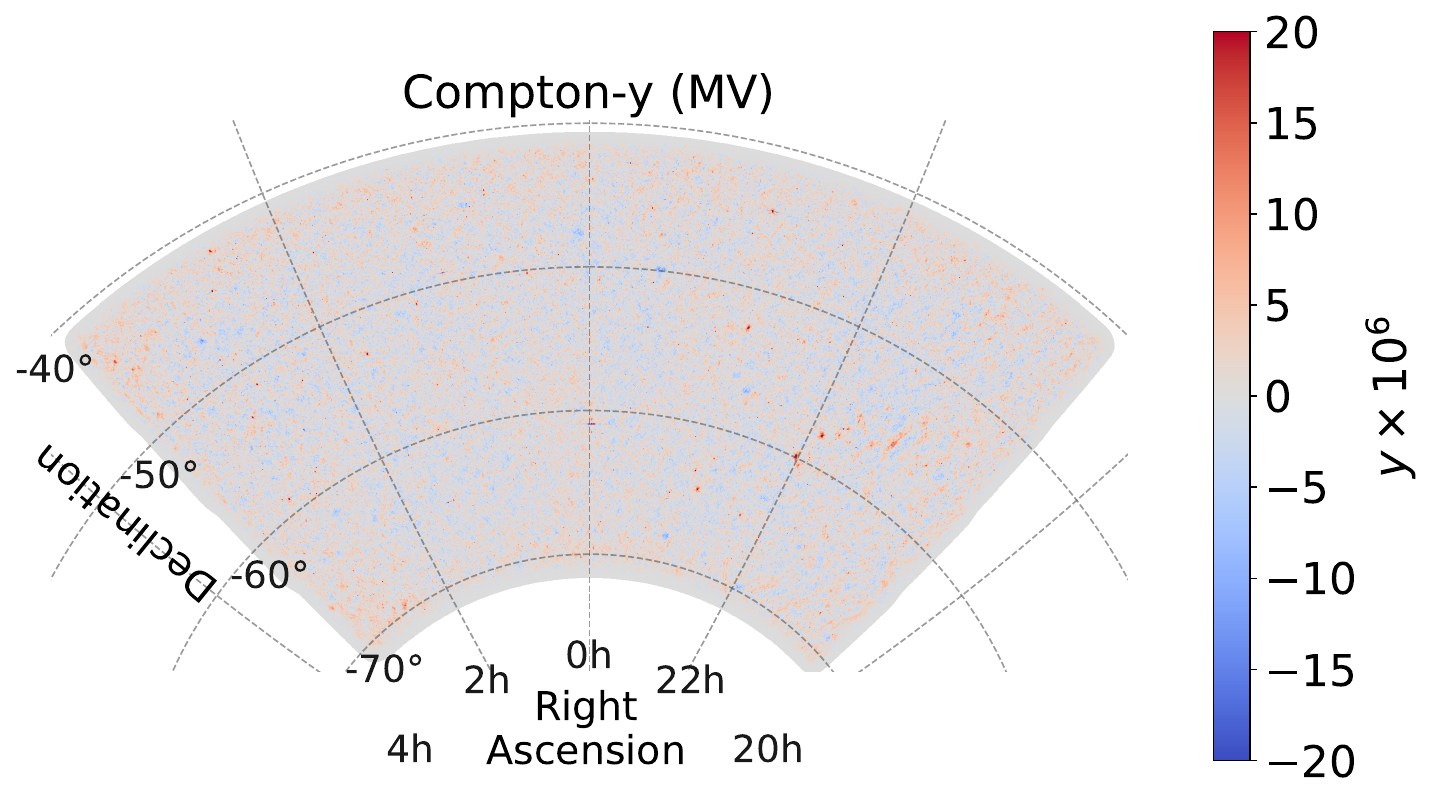}
    \caption{The \planck and SPT-3G combined minimum-variance (MV) version of the Compton-$y$ map provided with this work. We can see numerous galaxy clusters prominently visible as reddish features, resulting from the tSZ effect. }
    \label{fig:ymvplot}
\end{figure*}

\section{Resulting maps and their properties}
\label{sec:results}
\subsection{Maps}
We produce and release several versions of the Compton-$y$ maps: minimum-variance, CMB-deprojected, and CIB-deprojected. We provide these maps with $N_{\rm side} = 8192$, going up to $\ell = 13000$ with a ${\rm FWHM} = 1'.4$ beam. Our maps are dominated by \planck data on the largest angular scales where SPT-3G maps do not have useful information and vice-versa. We show the MV version of the Compton-$y$ map provided with this work in Fig.~\ref{fig:ymvplot}. Due to the high resolution and highly sensitive nature of our data, we can already point out numerous bright reddish point-like objects which are the galaxy clusters showing up in these maps due to the thermal SZ effect. This demonstrates the impressive depth of our Compton-$y$ maps. For each of our fiducial map versions (e.g., MV, CIB-deprojected), we also produce two corresponding `split' maps. These are generated by running our LC pipeline independently on two input data splits: (1) the first half of SPT data with the \planck even-ring map, and (2) the second half of SPT data with the \planck odd-ring map. The resulting pair of $y$-maps contains the same signal but has uncorrelated noise. In the next section, we discuss various properties of our maps using different statistical tools like power spectrum analysis and stacking.

\subsection{Properties of the Compton-$y$ maps}
\subsubsection{Power Spectrum Measurements}
\label{subsub:cibcross}
\begin{figure*}

    \begin{minipage}{0.5\textwidth}
        \centering
        \includegraphics[width=\textwidth]{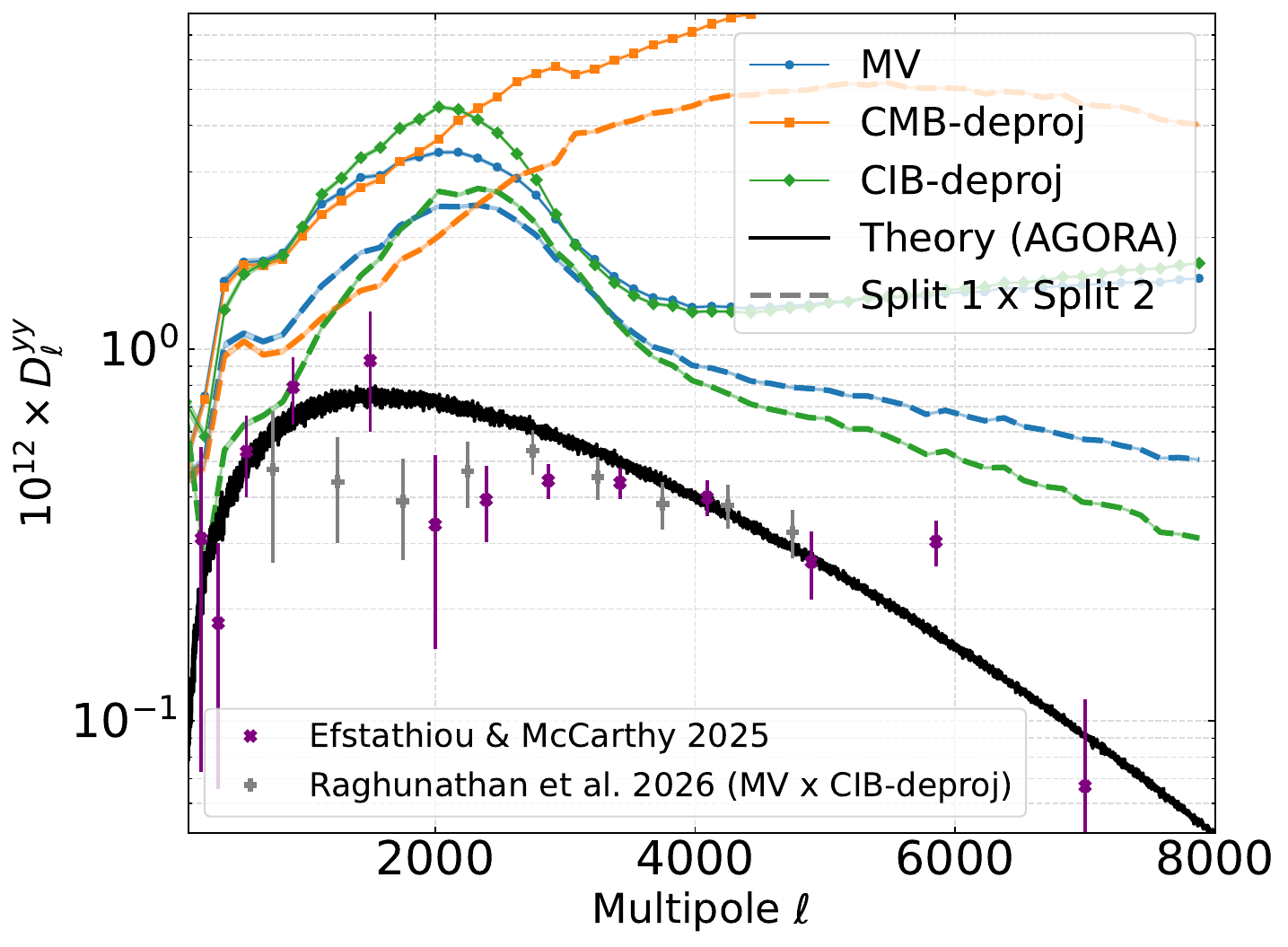}
    \end{minipage}\hfill
    \begin{minipage}{0.5\textwidth}
        \centering
        \includegraphics[width=\textwidth]{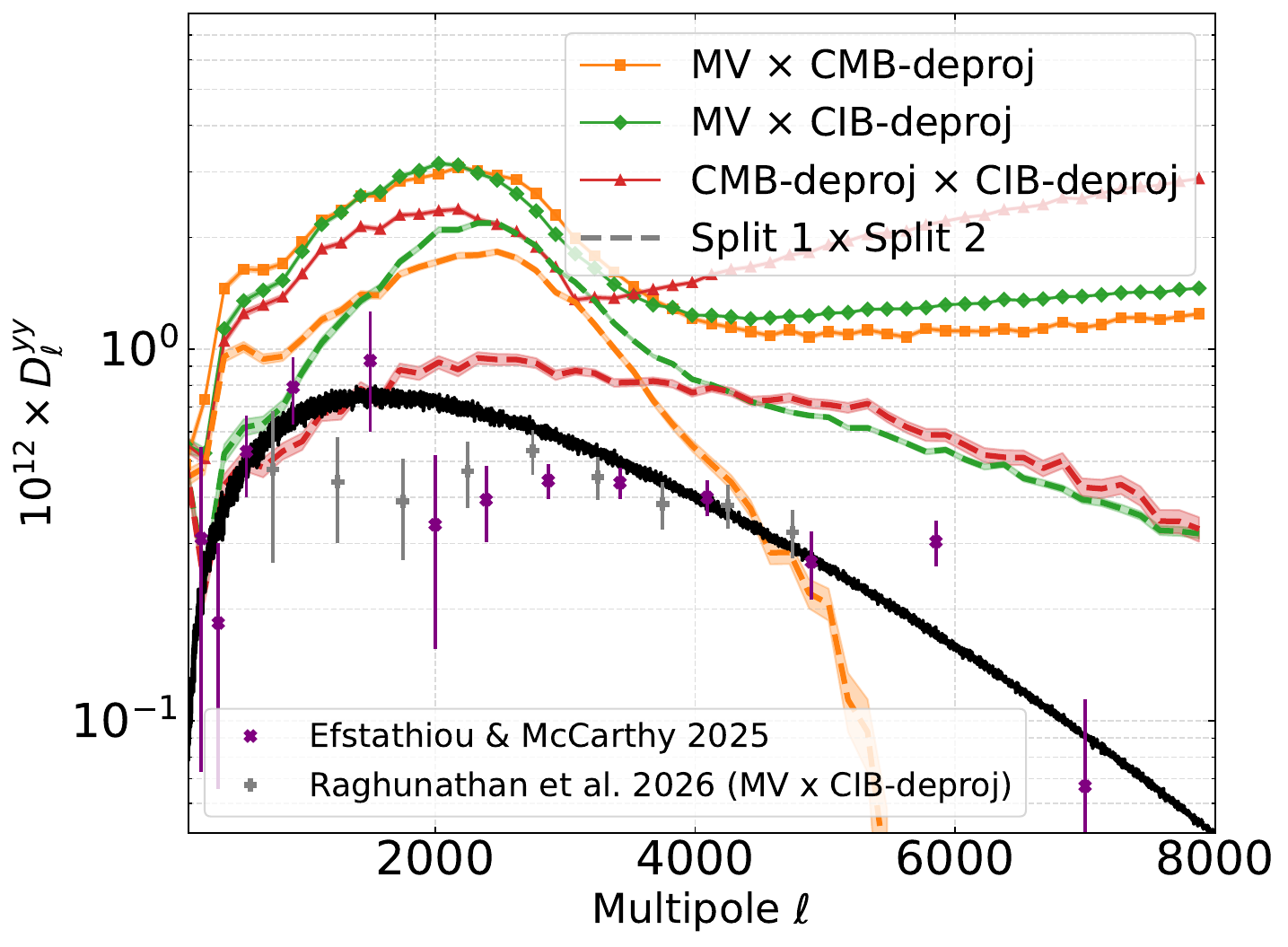} 
    \end{minipage}
 \caption{Auto-power spectra of different Compton-$y$ maps constructed in this work (left panel). The $x$-axis has been slightly offset in the left panel to aid the eye in comparing different curves. Cross-power spectra between combinations of the Compton-$y$ maps (right panel). Note that all the curves lie above the theory curve from \agora ~simulations because the maps are noise- and foreground-dominated at most scales (see Sec.~\ref{subsub:cibcross}). Here $D_\ell = \frac{\ell (\ell + 1)}{2\pi} C_\ell$. Shaded bands represent the Gaussian error bars on the power spectra. The black solid curve is the Compton-$y$ power spectrum from \agora ~simulations. Dashed curves are the corresponding split auto- and cross-spectra with two data splits to remove the noise bias and isolate the signal and residual foreground components. We also show measurements of the Compton-$y$ power spectrum from \cite{Efstathiou2025, Raghunathan2026} as purple and grey points respectively for comparison.}
\label{fig:clyy}
\end{figure*}

In the left panel of Fig.~\ref{fig:clyy}, we show auto-power spectra of different Compton-$y$ maps constructed in this work. These were computed using the \texttt{NaMaster} package \cite{Alonso2019}. We used a bin width of $\Delta_\ell = 150$ and the corresponding error bars (shown as shaded regions around the curves) are calculated analytically using the Gaussian approximation. \cite{Osato2021} show that non-Gaussian contributions to the covariance can be significant for Compton-$y$ power spectra while doing a cosmological analysis. But since we are not doing any parameter inference, we do not include such effects here. The Compton-$y$ maps are dominated by the combination of noise and residual foregrounds at most of the scales. As a result, all the measured power spectra lie above the theoretical predictions from \agora~simulations and estimated Compton-$y$ power spectrum by \cite{Efstathiou2025} using data from {\it Planck}, ACT, and SPT shown in purple points. 

By definition, the MV Compton-$y$ map has the least variance compared to other $y$ maps. So it might be slightly counterintuitive to observe that for certain scales, the power spectrum for the MV case is almost equal to or higher than the case where CMB or CIB are deprojected. This is especially visible for the case of CMB- and CIB-deprojected $y$ maps where their power spectra  lie below the MV case for some of the multipoles for $\ell \lesssim 1000$. This is a result of calculating weights using covariance matrices derived from simulations, where the optimal trade-off between noise and component leakage can differ slightly from the real data. Specifically, the simulation-optimized MV weights may allow a small leakage of dominant components that results in excess variance when applied to the real sky. In contrast, explicitly projecting out a CMB-like component drastically reduces the residual CMB and kSZ contamination in the resulting maps regardless of the covariance properties. Similarly, projecting out a modified blackbody SED helps in reducing the CIB and Galactic dust residuals from the final $y$ maps, but this is an imperfect operation since a modified blackbody is an approximation to the real SED for the CIB. At the same time, other components like extragalactic radio sources and the tSZ signal itself are correlated with the CIB. These correlations also contribute to the final power spectrum and an imperfect CIB modeling can potentially affect them as well resulting in biased power spectra on certain scales.

\begin{figure}
	\includegraphics[width=\columnwidth]{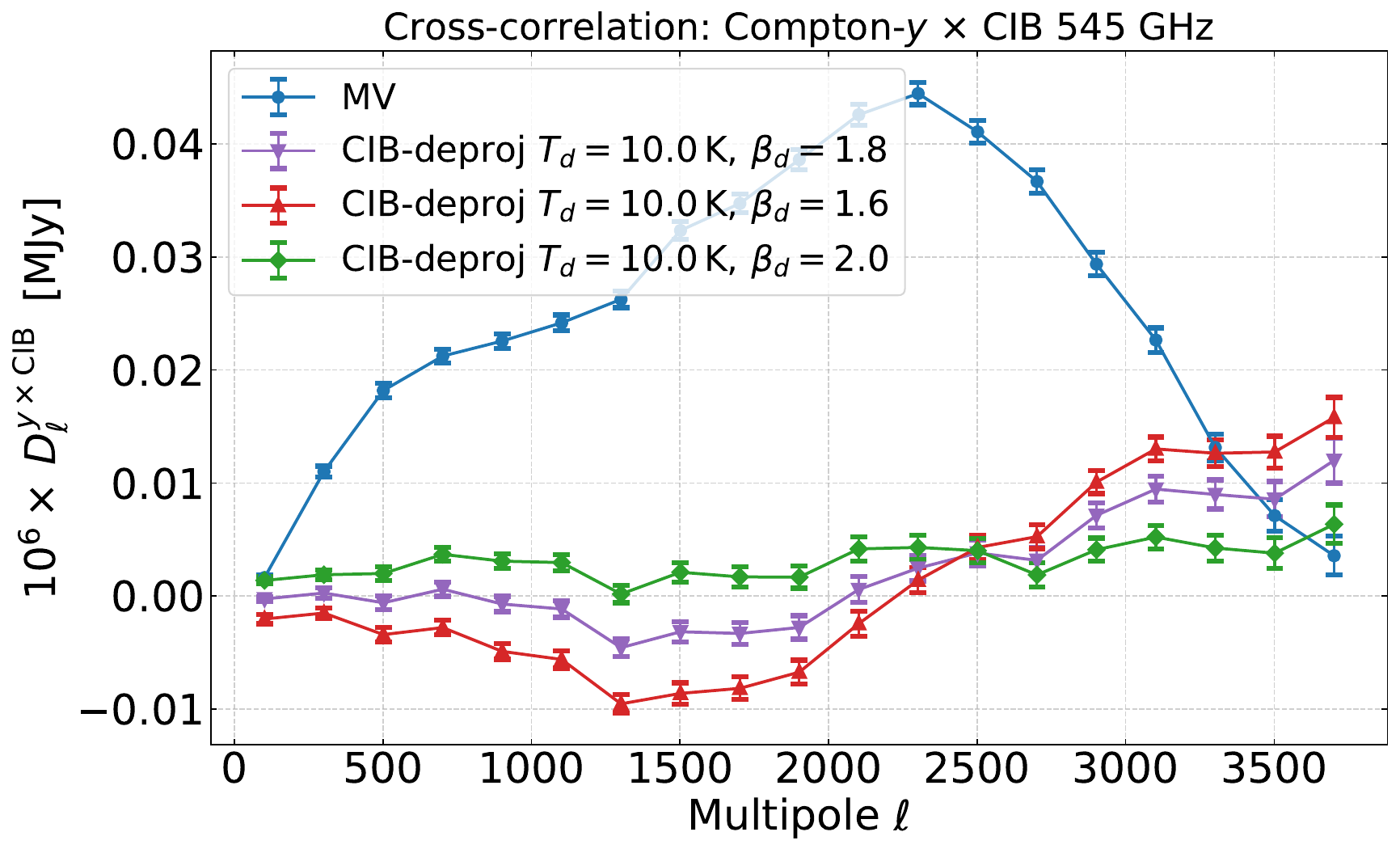}
    \caption{Cross-power spectra between our MV (blue) and CIB-deprojected Compton-$y$ maps (green, purple, and red)  and 545 GHz CIB maps from \cite{Lenz2019}. Different colors for the CIB-deprojected $y$ maps correspond to different dust emissivity index $\beta_d$ considered for the CIB while deprojecting it in our LC algorithm. The error bars are computed within the Gaussian approximation. While we see a positive correlation for the MV $y$ map, correlation with the CIB-deprojected maps is smaller and can vary based on the $\beta_d$ value. While this  points to potentially low level of CIB residuals in our CIB-deprojected Compton-$y$ maps, the differences in the residuals show the complex nature of the CIB and difficulties in modeling it accurately.}
    \label{fig:y545cross}
\end{figure}

For CIB residuals, this can be further probed by cross-correlating the resulting $y$ maps with other tracers of the large-scale structure. For example, we would expect a lower amplitude for the CIB-deprojected $y$ map compared to the MV $y$ map when cross-correlated with a CIB map. This is exactly what we see in Fig.~\ref{fig:y545cross} where we use the CIB map at \planck 545 GHz frequency channel from \cite{Lenz2019} (with error bars calculated using the Gaussian approximation). This map is constructed using the \planck HFI intensity maps at 353, 545, and 857 GHz and cleaning out the galactic dust using an external template for neutral atomic hydrogen which is an excellent tracer of the galactic dust. This procedure is carried out over 25\% of the sky in relatively cleaner (lesser galactic dust contamination) patches of the sky near the northern and southern galactic poles. The common sky fraction between this map and our Compton-$y$ maps is $\sim 2.6\%$. We calculate the cross-power spectrum between this map and our MV and CIB-deprojected maps as shown in Fig.~\ref{fig:y545cross}. In particular, we show the cross-correlation for three different CIB-deprojected Compton-$y$ maps with each having the same $T_d = 10.0$ K parameter but different $\beta_d$ values ($\beta_d = 1.6, 1.8, 2.0$ in red, purple, and green color respectively). While we definitely see a positive correlation between the MV $y$ map and the CIB 545 GHz map, the correlation with the CIB-deprojected $y$ map is much smaller comparatively. While \cite{Lenz2019} do not explicitly remove any tSZ contamination from their maps, at 545 GHz, the CIB is expected to be much more dominant than the tSZ effect. Our cross-correlation spectra are therefore expected to be dominated by this cross-correlation between the CIB in 545 GHz maps and tSZ signal in our Compton-$y$ maps. We repeat the same exercise with the CIB maps at 857 GHz (where the CIB is even more dominant than tSZ signal) and find the resulting CIB-deprojected $y$ map cross-spectra consistent with the 545 GHz CIB map and with the findings of \cite{McCarthy2024}. This result confirms that our CIB-deprojected maps are much less contaminated by the residual CIB than the MV Compton-$y$ map. 

At the same time, it shows the difficulty in modeling the CIB contribution. The cross-correlation with the 545~GHz CIB map seems to be significantly different depending on the parameters used for the CIB SED. In fact, for $T_d = 10$ K and $\beta_d=1.6$, the correlation is negative for $\ell \lesssim 2000$. These results are due to the complex interplay of how different \planck and SPT frequency maps are weighted before being added to get the final $y$ map. We test this on \agora~simulations as shown in Fig.~\ref{fig:y545cross_agora} (see App.~\ref{app:betadust_discus} for a discussion) and make similar observations that the CIB residuals vary with different $T_d$ and $\beta_d$ combinations. Going forward, for the CIB-deprojected maps, we only show results with the $T_d=10$ K and $\beta_d = 2.0$ parameter values. 

The dashed curves in the left panel of Fig.~\ref{fig:clyy} show the cross-power spectrum of the two independent data splits for each map. This cross-spectrum is free of the noise bias and thus isolates the contribution from the true signal and foregrounds. While the auto-power spectrum of the full CIB-deprojected map is equal to or higher than the MV case at $\ell > 3500$, the CIB-deprojected split-power spectrum is lower than the MV case. This is due to the increased noise penalty from the LC algorithm when we deproject a component. 

\subsubsection{Compton-$y$ Power Spectrum Estimation}
Our Compton-$y$ maps have residual foregrounds in them including the CIB, galactic and extragalactic radio emission, CMB, kSZ etc. Some of these foregrounds (CIB and extragalactic radio emission) are correlated with the tSZ effect since they are all biased tracers of the large-scale structure. On top of that, the CIB and radio emission are also expected to be correlated. As a result, measured power spectra of these Compton-$y$ maps are a sum of the residual noise and correlated and uncorrelated foregrounds. The measured power spectrum therefore is not an unbiased estimator of the true tSZ power spectrum. 

We can use constrained LC techniques to significantly reduce the contamination from one or more foregrounds. As a result, combinations of Compton-y maps with different foreground mitigation can potentially yield less biased measurement of the tSZ power spectrum. For example, since the CMB and CIB are the two most significant foregrounds to the tSZ signal, we could cross-correlate the CMB- and CIB-deprojected $y$ maps to get a potentially unbiased estimation of the tSZ power spectrum. This is what we show in the solid red curve in the right panel of Fig.~\ref{fig:clyy}. We can see that the cross-power spectrum is lower than the corresponding auto-power spectra. This potentially means that the residual foreground biases have been suppressed better than in the auto-power spectrum case. Similar observations can also be made for the other combinations of the $y$ maps plotted in the right panel of Fig.~\ref{fig:clyy}. 

However, caution has to be exercised while interpreting these measurements. 
It is important to recognize that these cross-combinations of $y$-maps, while effective at mitigating certain major contaminants, can introduce smaller sources of bias. For instance, a $y$-map that deprojects the CIB may suffer from an increased CMB and radio contamination level, while a CMB-deprojected $y$-map may retain stronger contamination from radio galaxies and the CIB. Since both the CIB and radio emission trace large-scale structure, their residual contributions can remain correlated with the true tSZ signal, thereby biasing the recovered power spectrum. As discussed in Sec.~\ref{subsub:cibcross}, the amplitude of this bias can also change slightly depending on the choices of parameters of the SED to remove the CIB. 

In addition, deprojection alters the noise properties of the maps, which complicates the interpretation of the cross-spectra due to correlated noise between these maps. This means their cross-power spectrum (solid lines in the right panel of Fig.~\ref{fig:clyy}) is dominated by a large noise bias. We mitigate this by instead cross-correlating the noise-uncorrelated splits (e.g., split 1 of the CMB-deprojected map with split 2 of the CIB-deprojected map). This removes the noise bias, as shown by the dashed curves in the right panel of Fig.~\ref{fig:clyy}. This `split cross-spectrum' leaves us with an estimate of the tSZ signal, contaminated only by relatively smaller, correlated foreground biases discussed above.

With that in mind, we conclude that the CMB-deprojected $\times$ CIB-deprojected split cross-spectrum is the most robust product for estimating the true tSZ power spectrum while accounting for residual foregrounds. It effectively suppresses the two largest contaminants (CMB and CIB) and our use of splits removes the noise bias. For comparison, we plot external measurements from \cite{Efstathiou2025} and \cite{Raghunathan2026}. 
Instead of relying on the measurements from a Compton-$y$ map, \cite{Efstathiou2025} estimate the tSZ power spectrum by a combination of data and models/templates applied to the frequency maps from {\it Planck}, SPT, and ACT to account for the non-tSZ components. On the other hand, \cite{Raghunathan2026} combine the data from SPT and \textit{Herschel}-SPIRE surveys over 100 ${\rm deg}^2$ and use the LC technique to construct different $y$ maps (MV, CIB-deprojected etc). They account for the residual foreground power in their maps through an estimate from \agora~simulations. We have not subtracted an estimate of the residual foregrounds in our power spectra in Fig.~\ref{fig:clyy} and thus do not expect our curves to match with data. However, the comparison between the auto- and cross- split power spectrum 
shows that this combination, which suppresses the two most significant contaminants, is a promising pathway to a robust tSZ power spectrum measurement. Similarly, our MV $\times$ CIB-deprojected split cross-spectrum at $\ell > 4000$ can be used as another potential probe of the Compton-$y$ power spectrum with a final careful accounting of the remaining foreground residuals still required. 

One potential approach to interpret the corresponding tSZ and residual foreground contamination is to start by fitting realistic foreground models/templates to the frequency auto- and cross-power spectra, and then to propagate the resulting signal and foreground best-fit models/templates through our 2D LC weights to arrive at the expected tSZ and foreground contamination values. This will be discussed in more detail in an upcoming paper (Maniyar \textit{et al.}, in preparation).

The MV $\times$ CMB-deprojected split cross-spectrum drops off significantly at $\ell \gtrsim 2500$. This happens due to a strong negative correlation between the residual CIB in the two maps. The CIB contamination in the MV maps drops after $\ell > 2500$ which can be seen from the drop in the split auto-spectrum of the MV map (blue dashed curve in the left panel of Fig.~\ref{fig:clyy}). It can also be seen from the drop in the cross-power spectrum between the MV map and the CIB map at 545 GHz in Fig.~\ref{fig:y545cross} (blue curve) at $\ell > 2500$. On the other hand, the split auto-spectrum of the CMB-deprojected $y$ map rises at $\ell > 2500$ due to the increased CIB residuals (orange dashed curve in the left panel of Fig.~\ref{fig:clyy}). This behaviour results in a negative correlation bringing the MV $\times$ CMB-deprojected split cross-spectrum down. We observe a similar behavior in our simulated $y$ maps. As will be discussed in the next section, such behavior can also be interpreted from the respective drop and rise of the cross-correlation of the MV and CIB-deprojected Compton-$y$ maps with unWISE galaxies at $\ell \gtrsim 2500$.

\subsubsection{Foreground Inferences Through Cross-correlation With unWISE Galaxy Samples}
\label{sec:unwise_cross}

To further explore the astrophysical content of our $y$-maps, we cross-correlate them with galaxy density maps from the unWISE catalog \cite{Schlafly_2019, Krolewski2020}. The unWISE selection provides photometric galaxy samples with large sky coverage and broad redshift distributions. Here, we focus on two subsamples: the `blue' sample, dominated by relatively lower-redshift galaxies with median $z \sim 0.6$, and the `red' sample, which peaks at somewhat higher redshift ($z \sim 1.1$). These two samples are complementary probes of the large-scale structure at different epochs.

\begin{figure}
	\includegraphics[width=\columnwidth]{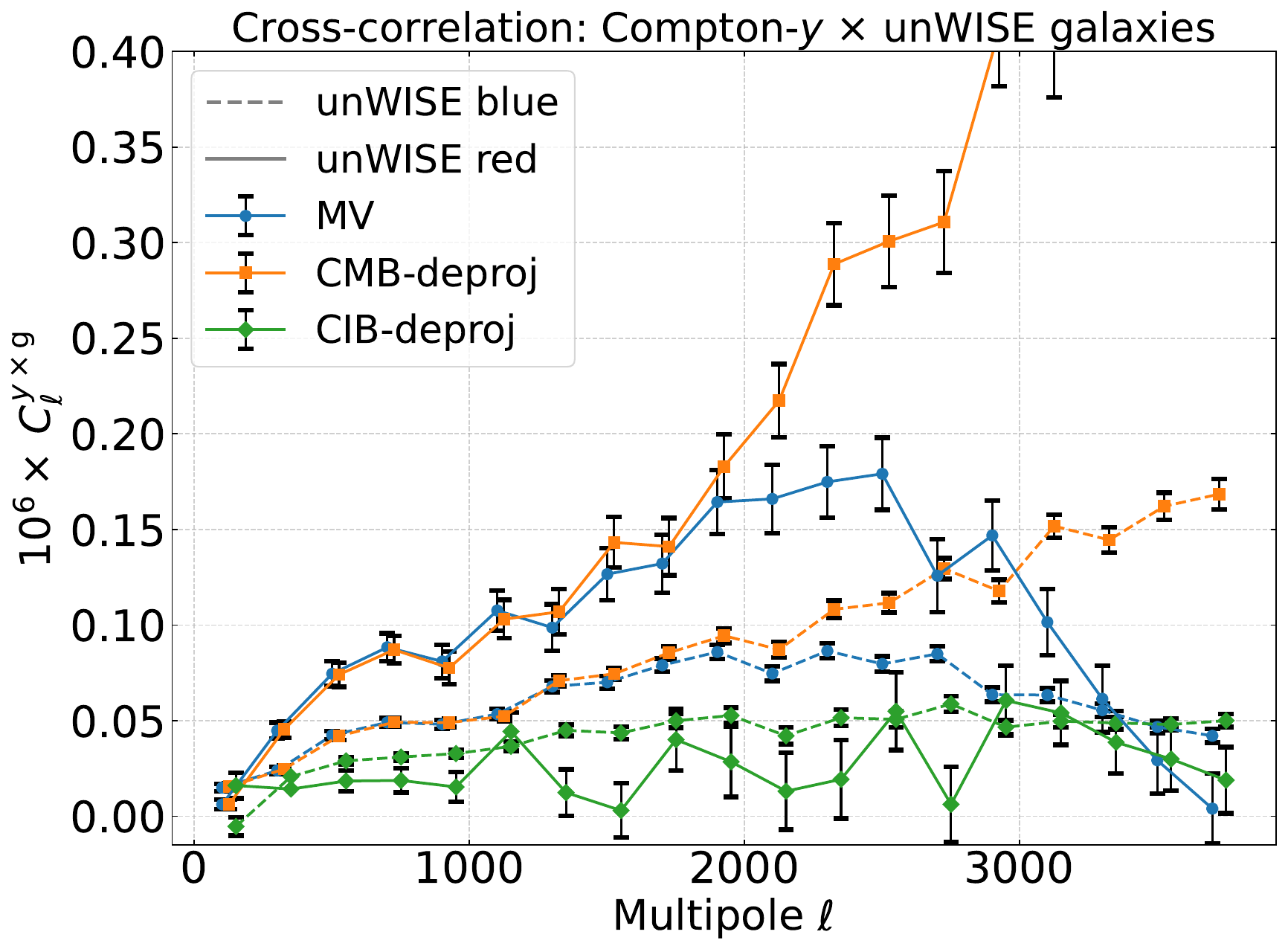}
    \caption{Cross-correlation between the Compton-$y$ maps from this work and the unWISE blue and red galaxy samples in dashed and solid lines respectively. The error bars are calculated within the Gaussian approximation. The red sample peaks at higher redshift ($z \sim 1$) than the blue sample which predominantly traces galaxies at relatively low redshift ($z \sim 0.6$). 
    Different colors show results for the MV, CMB-deprojected, and CIB-deprojected $y$-maps. The lower correlation observed for the CIB-deprojected maps than MV and CMB-deprojected maps, especially for the red galaxy sample at higher redshift where CIB contributes strongly, is consistent with suppressed CIB residuals in the former maps. This result is a preview of an upcoming analysis from Maniyar \textit{et al.} (in preparation).  
    }
    \label{fig:unwise_both}
\end{figure}

Figure~\ref{fig:unwise_both} shows the cross-correlation between our Compton-$y$ maps and the blue (dashed lines) and red (solid lines) unWISE galaxy samples. The results point to the different physical origins of the $y$-map signal and its foregrounds. As shown in \cite{Komatsu2002, Bhattacharya2012, Maniyar2023}, a majority of the contribution to the tSZ effect is sourced by hot electrons residing in massive halos ($M_{\rm halo} \gtrsim 10^{14} M_\odot$) at relatively low redshifts ($z \lesssim 1$), where clusters and groups contribute most of the signal. In contrast, the CIB originates from dusty, star-forming galaxies hosted by lower-mass halos ($M_{\rm halo} \sim 10^{11} - 10^{13} M_\odot$), which are distributed over a broader redshift range but peak around $z \sim 1$--$2$ (\cite{Maniyar2021, Yan2024}). We find that the CIB-deprojected $y$-maps exhibit a positive correlation with the blue sample, while their correlation with the red sample is smaller. On the other hand, the MV and CMB-deprojected maps, which retain more CIB contamination, show a stronger correlation with the red sample. This suggests a trend of higher tSZ contribution from low-redshift, massive halos in the blue sample and CIB-dominated correlation in the red sample. These trends demonstrate that such cross-correlations, through the redshift dependence of the galaxy sample, combined with the differing mass and redshift contributions of tSZ signal and CIB, provide useful tests of residual foreground contamination. These results provide a clear diagnostic. The CIB-deprojected map is the recommended product for any cross-correlation analysis with other large-scale structure tracers where CIB contamination is a concern (e.g., CMB lensing, high-redshift galaxy catalogs, or other CIB tracers). 

It is important to note that such cross-correlations must be interpreted by keeping residual foregrounds in mind as discussed in Sec.~\ref{subsub:cibcross}. The unWISE galaxy samples are themselves biased tracers of the underlying matter distribution, and their overlap with residual foregrounds can bias the cross-correlation measurement. For example, radio galaxy residuals in the Compton-$y$ maps can contribute to these correlations. In addition, uncertainties in the photometric redshift distributions of the blue and red samples complicate the interpretation of the exact redshift dependence. For such reasons, while the observed trends are qualitatively consistent with expectations for tSZ and CIB, a careful treatment of systematic biases is required before drawing quantitative conclusions. 
In an upcoming paper (Maniyar \textit{et al.}, in preparation), we present a detailed analysis of this cross-correlation with unWISE galaxies. 

\subsubsection{Stacked Aperture Photometry Profiles at Cluster Locations}
\label{subsub:ap_stacking}

To further characterize the tSZ signal in our Compton-$y$ maps, we perform a stacked aperture photometry (AP) analysis at the positions of clusters (total 4568 clusters detected with a signal-to-noise greater than five) from the five-year SPT-3G cluster sample (Bleem \textit{et al.}, in preparation) with a median redshift of $z = 0.71$. We measure the mean signal within circular apertures of increasing radius and subtract the mean signal in a surrounding annulus, constructing an AP profile as a function of radius. This method provides an intuitive measure of the integrated cluster tSZ profile while suppressing background fluctuations. 

To mitigate residual scan-aligned artifacts that produce faint horizontal `wings' in the stacked Compton-$y$ maps (see \cite{Quan2026, Camphuis2025, Kayla2025} for more details), we applied a simple baseline correction to each cutout prior to stacking. Although the inclusion of Planck data restores the largest-scale modes, these small-scale scan-direction offsets persist at high multipoles where the reconstruction is dominated by SPT. For each stamp, we subtracted the median value computed along each map row, excluding a circular region of radius 2' centered on the cluster to preserve the compact signal. This procedure effectively removes large-scale striping or scanning offsets introduced by the map-making process while leaving the central tSZ profile unaffected. \cite{Kayla2025} employ a more sophisticated approach for their SPT-only analysis, constructing an empirical point-source template from median-stacked, scan-aligned cutouts of bright sources that is then scaled and iteratively subtracted from the maps before masking the brightest residual sources. As our focus here is on demonstrating the properties of the constructed Compton-$y$ maps, we do not implement the full template-subtraction procedure of \cite{Kayla2025} and instead apply this simplified correction.

\begin{figure}
    \centering
    \includegraphics[width=\columnwidth]{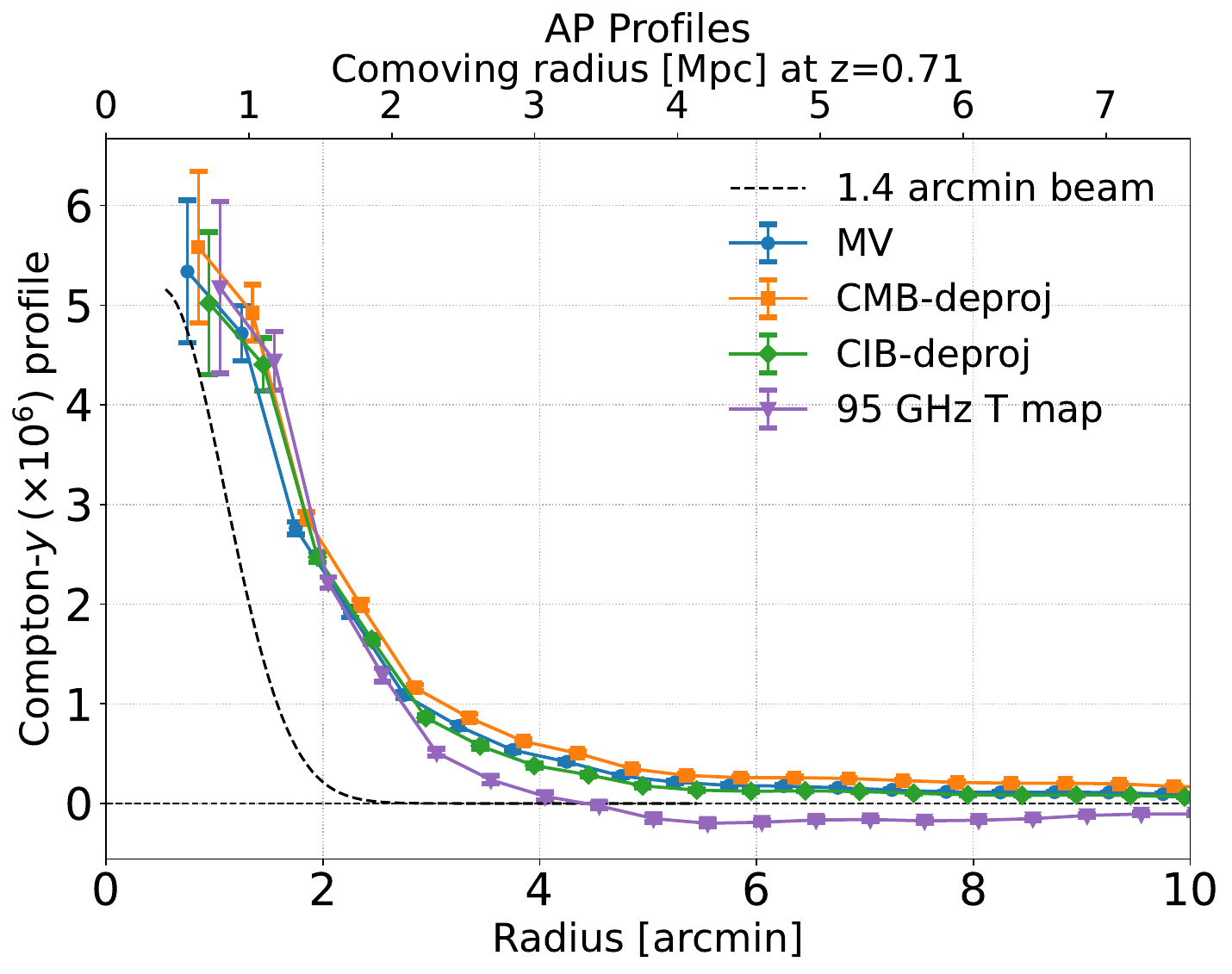} 
    \caption{Aperture photometry profiles for cluster stacks across different Compton-$y$ maps. The MV (blue), CMB-deprojected (orange), and CIB-deprojected (green) Compton-$y$ maps remain positive across all radii showing negligible CIB contamination at cluster locations. The 95\,GHz map (purple) shows a negative tail beyond $\sim4'$ due to the missing low-$m$ and low-$\ell$ modes in the SPT-3G map-making.  Error bars denote the pixel variance within each annulus. The profiles have been plotted with a slight horizontal offset to aid visual inspection. We also show a $1'.4$ Gaussian beam, overplotted for visual reference (black dashed curve).}
    \label{fig:ap_full}
\end{figure}

Figure~\ref{fig:ap_full} shows the radial profiles for our Compton-$y$ maps. We also show the radial profile for SPT 95\,GHz temperature map converted to Compton-$y$ units using Eq.~\ref{eq:ysed}. The 95\,GHz temperature map exhibits a negative tail at radii beyond $\sim4'$, reflecting the impact of the instrumental transfer function and high-pass filtering ($\ell < 300$) used during SPT-3G map-making. 
In contrast, the $y$ maps remain positive across all radii, indicating that the inclusion of Planck data effectively restores these missing modes. 
We leave a detailed investigation of these profiles for a future study. 

\begin{figure}
    \centering
    \includegraphics[width=\columnwidth]{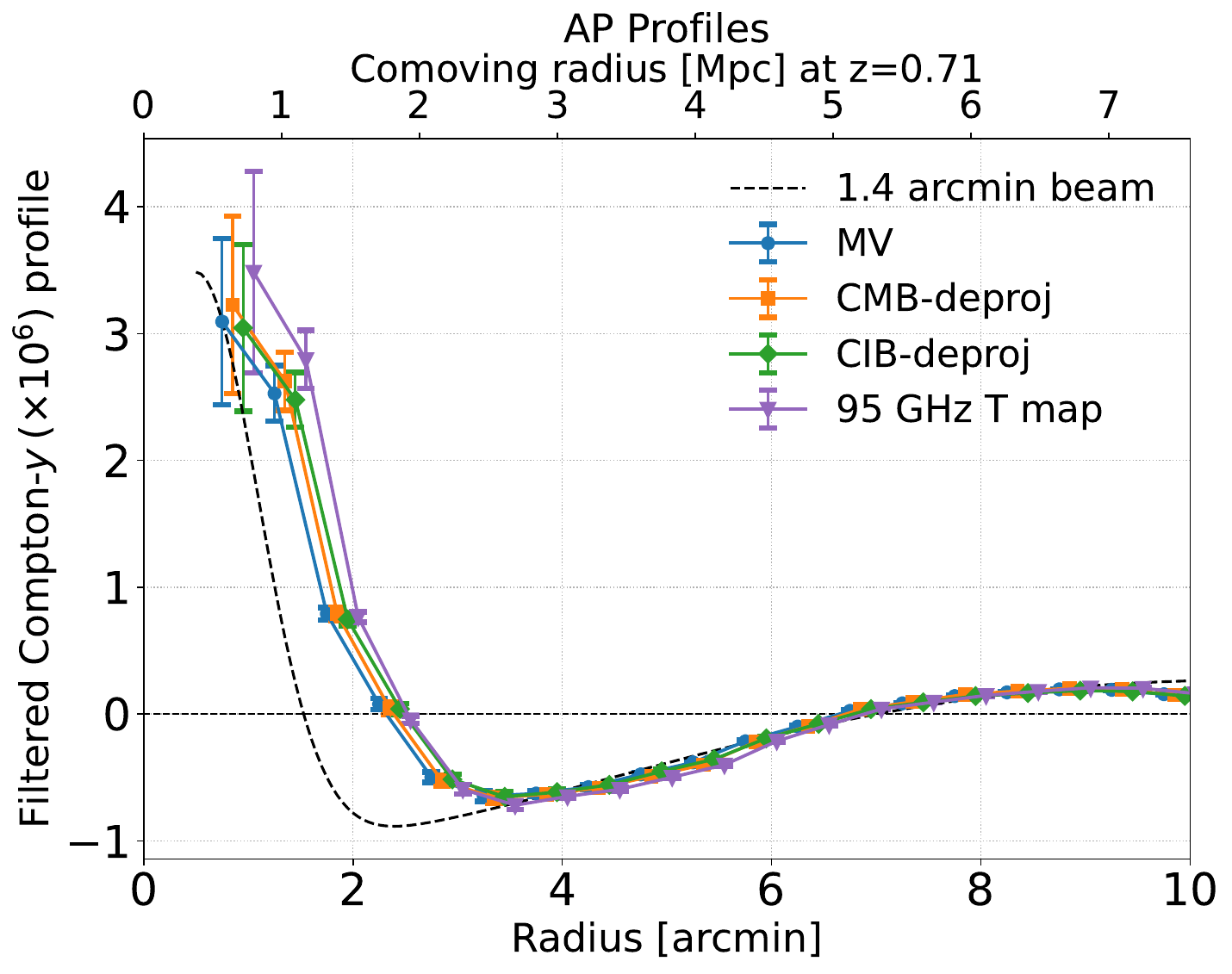} 
    \caption{AP profiles for the same cluster stacks after high-pass filtering the maps with a smooth cutoff starting at $\ell=2000$. Filtering removes large-scale information, yielding consistent profiles across all maps. The close agreement among the profiles highlights the robustness of the recovered tSZ signal and its insensitivity to the precise foreground-deprojection scheme. The profiles have been plotted with a slight horizontal offset to aid visual inspection. A $1'.4$ Gaussian beam, high-pass filtered in the same manner as the data, is overplotted for visual reference (black dashed). }
    \label{fig:ap_hpf}
\end{figure}

To isolate the cluster tSZ signal on small scales, we apply a smooth harmonic-space high-pass filter starting at $\ell>2000$ to our Compton-$y$ maps and repeat the stacking exercise. The high-pass filtering step ($\ell > 2000$) removes the large-scale modes that distinguish the $y$-maps from the SPT-only maps. This choice isolates the small-scale tSZ signal and other residuals, while suppressing spurious large-scale fluctuations. 
The resulting profiles, shown in Figure~\ref{fig:ap_hpf}, demonstrate that all maps, including the 95\,GHz map, trace the same small-scale tSZ signal. 
These results highlight the robustness of the recovered tSZ signal in our Compton-$y$ maps and its consistency with the high-resolution SPT-3G data. The comparison to the $95\,$GHz temperature map is instructive. We filter the 95 GHz temperature map (converted to Compton-$y$ units) in the same manner. The stacked cutouts then yield a profile that traces the same small-scale structure, as we would expect, demonstrating that the $95\,$GHz channel contains a substantial tSZ signal. This result confirms the findings in \cite{Kayla2025} who also observed robustness of the tSZ signal in SPT data at cluster locations. 

Beyond the illustrative comparison shown here, the combined analysis of full and high-pass filtered profiles establishes a promising avenue for further astrophysical inference. In particular, they can be used to constrain average intracluster pressure profiles, test for deviations from the expected virialized form, or probe mass and redshift trends in cluster populations. While such an analysis is beyond the scope of this work, the consistency of the stacked profiles across different map constructions provides a foundation for these future studies. Our analysis shows that for objects like galaxy clusters, the tSZ signal dominates over foreground residuals on small scales. All maps produce a consistent profile. For such science cases (e.g., cluster pressure profile fitting), when the dominant uncertainty is the total variance noise, we recommend using the MV Compton-$y$ map.

\section{Conclusions and Future Prospects} \label{sec:conclusion}

In this work we constructed a new suite of thermal Sunyaev-Zel’dovich Compton-$y$ maps from the combination of SPT-3G D1 data over $1500\,{\rm deg}^2$ with the \textit{Planck} PR3 all-sky data. Our analysis employs a harmonic-space linear combination framework with $(\ell,m)$-dependent weights, producing multiple reconstructions designed to balance statistical sensitivity against robustness to astrophysical contaminants. In particular, we present minimum-variance, CMB-deprojected, and CIB-deprojected maps. This set of complementary products provides flexibility for a broad range of astrophysical and cosmological applications.

We validated these maps through a series of statistical tests. The angular power spectra highlight the expected trade-offs between noise and foreground removal, with the CIB-deprojected maps exhibiting suppressed correlations with an external CIB map compared to the MV map. Cross-correlations with unWISE galaxy samples reveal trends consistent with the differing redshift and halo-mass contributions of the tSZ effect and the CIB, further supporting the robustness of the reconstructed $y$ signal. We also perform a stacking and aperture photometry analysis at SPT-3G cluster locations which provide an illustration of the tSZ profiles. We observe that the small-scale stacked profiles converge to consistent and robust measurements of the cluster tSZ signal. Together, these results demonstrate that different Compton-$y$ maps presented here can be used for several scientific explorations depending on the noise vs. residual bias concerns.

An important caveat concerns the treatment of the CIB. While our deprojection methods are effective at suppressing CIB contamination, the complex SED of the CIB makes it difficult to model with a single modified blackbody template. In particular, we find that the choice of dust emissivity index $\beta_{d}$ can affect the level of residual contamination, leading to variations in the measured cross-correlations with external tracers. Our results highlight that our CIB-deprojected maps provide a substantial reduction of foreground bias, but they do not guarantee a completely unbiased tSZ reconstruction. Therefore, careful foreground modeling should be exercised when employing these maps for analyses where residual CIB contamination could play a significant role compared to the precision of the data being considered.

We will make the data products from this work publicly available: full and split Compton-$y$ maps (MV, CMB-deprojected, CIB-deprojected with $T_d=10.0 \, {\rm K}, \, \beta_d=2.0$) at $N_{\rm side} = 8192$ with a beam of FWHM of 1'.4 and $\ell_{\rm max}=13000$, the harmonic space transfer function to account for the missing modes (Sec.~\ref{subsub:mmodesptplanck}), and a point source catalog (Archipley \textit{et al.}, in preparation) from the five-year SPT-3G data (2019-2024) down to 6 mJy, upon acceptance of the paper. These maps represent the highest-resolution Compton-$y$ maps currently available over the SPT-3G survey region. They open the door to several immediate scientific applications, including precise power spectrum and bispectrum measurements and analyses, studies of cluster pressure profiles, and cross-correlation analyses with large-scale structure tracers from optical to infrared surveys. The cross-correlation studies are possible due to the overlap of the SPT-3G footprint with existing and upcoming deep imaging data sets such as the Dark Energy Survey \cite{des2016}, \textit{Euclid} \cite{Euclid2025}, and SPHEREx \cite{spherex2020} mission, enabling new constraints on the interplay between galaxies, hot gas, and cosmological parameters. 
Based on our validation tests:
(1) For tSZ power spectrum estimation, we recommend using the CMB-deprojected $\times$ CIB-deprojected split cross-spectrum (while accounting for residual foregrounds), which suppresses the major foregrounds and removes noise bias.
(2) For cross-correlations with large-scale structure tracers (especially high-redshift or CIB-like catalogs), we recommend the CIB-deprojected map, which has suppressed CIB contamination.
(3) For individual object analysis (e.g., cluster pressure profiles), we recommend the MV map for the lowest total variance.

On longer timescales, SPT-3G data will deliver deeper and wider Compton-$y$ maps, and the methodology developed here will be readily extendable to upcoming experiments such as the Simons Observatory \cite{SO2019}, which promises highly sensitive data over a large fraction of the sky. These future data sets will allow for precision measurements of the tSZ effect across a wide dynamic range in redshift and mass, providing a comprehensive view of the thermal history of baryons.

\section{Acknowledgments}
\label{sec:acknowledgements}
The South Pole Telescope program is supported by the National Science Foundation (NSF) through awards OPP-1852617 and OPP-2332483. Partial support is also provided by the Kavli Institute of Cosmological Physics at the University of Chicago. 
Argonne National Laboratory’s work was supported by the U.S. Department of Energy, Office of High Energy Physics, under contract DE-AC02-06CH11357. 
The UC Davis group acknowledges support from Michael and Ester Vaida. 
Work at the Fermi National Accelerator Laboratory (Fermilab), a U.S. Department of Energy, Office of Science, Office of High Energy Physics HEP User Facility, is managed by Fermi Forward Discovery Group, LLC, acting under Contract No. 89243024CSC000002.
The Melbourne authors acknowledge support from the Australian Research Council’s Discovery Project scheme (No. DP210102386). 
The Paris group has received funding from the European Research Council (ERC) under the European Union’s Horizon 2020 research and innovation program (grant agreement No 101001897), and funding from the Centre National d’Etudes Spatiales. 
The SLAC group is supported in part by the Department of Energy at SLAC National Accelerator Laboratory, under contract DE-AC02-76SF00515.

This work used the following libraries for numerical, scientific computing, and plotting: \texttt{HEALPix}~\citep{healpix2005}, \texttt{NumPy}~\citep{numpy}, \texttt{SciPy}~\citep{scipy}, and \texttt{Matplotlib}~\citep{matplotlib}.

\appendix

\section{Effect of varying the $\beta_d$ parameter in simulations}
\begin{figure}
	\includegraphics[width=\columnwidth]{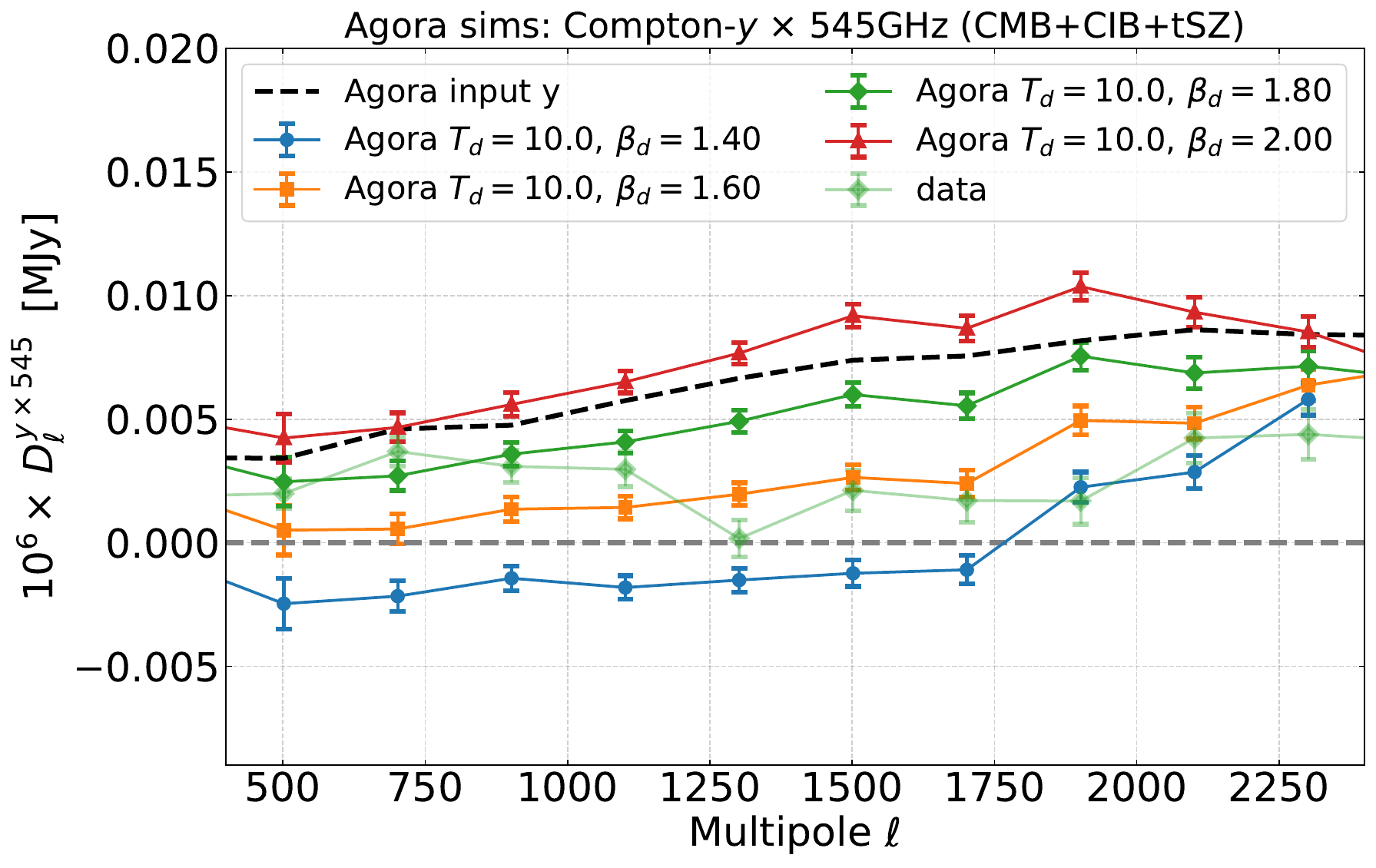}
    \caption{Cross-power spectra between several simulated CIB-deprojected Compton-$y$ maps and 545 GHz maps using \agora~simulations. Different colors for the CIB-deprojected $y$ maps correspond to different dust emissivity indices $\beta_d$ considered for the CIB while deprojecting it in our LC algorithm. The 545 GHz map is constructed by convolving the CMB, CIB, and tSZ maps from \agora~simulations with \planck 545 GHz bandpass filter and combining them. The black dashed curve is the cross-correlation between the input \agora~Compton-$y$ maps and simulated 545 GHz map. The error bars are computed within the Gaussian approximation. The cross-correlation levels change with different $\beta_d$ values. The faded green curve shows the measurement of our CIB-deprojected Compton-$y$ map and 545 GHz CIB map from \cite{Lenz2019}. This comparison shows the complex nature of the CIB and difficulties in modeling it accurately.}
    \label{fig:y545cross_agora}
\end{figure}

As mentioned in Sect.~\ref{subsub:freqresp}, the CIB SED is complex and hard to model analytically. In this work, we approximate the CIB SED with a modified blackbody with the dust temperature and emissivity parameters to be $T_d = 10$ K and $\beta_d = 2.00$. In Fig.~\ref{fig:y545cross}, we observed that the cross-correlation between our CIB-deprojected Compton-$y$ maps with 545 GHz CIB maps from \cite{Lenz2019} depends on the value assumed for the $\beta_d$ parameter for a given value of $T_d$. We see a similar trend when we test this on \agora~simulations as shown in Fig.~\ref{fig:y545cross_agora}. We simulate several CIB-deprojected Compton-$y$ maps with different $\beta_d$ values. We then cross-correlate these maps with a simulated 545 GHz frequency map containing the CMB, CIB, and tSZ. We see a similar pattern in simulations where this cross-correlation is sensitive to the value of $\beta_d$. The $T_d = 10$ K and $\beta_d = 2.00$ curve in Fig.~\ref{fig:y545cross_agora} from simulations differs quite a bit with the actual cross-correlation measured from data with the same parameter values, although both of them provide a positive correlation. This exercise goes on to further confirm the complexities involved in correctly modeling and simulating the CIB, and cleanly removing it from the maps. While the CIB-deprojection method suppresses the CIB contamination from the maps significantly, using these maps from cross-correlation studies requires considering the level of residual bias in the map. 

\label{app:betadust_discus}
\bibliography{bibfile}

\end{document}